# A frictional control mechanism of circumpolar transport in barotropic reentrant channel models

Takuro Matsuta[1], Atsushi Kubokawa[1], Humio Mitsudera[2] and Tomomichi Ogata[3].

[1] Faculty of Environmental Earth Science, Hokkaido University, Sapporo, Japan.

[2] Pan-Okhotsk Research Center, Institute of Low Temperature Science, Hokkaido University, Sapporo, Japan.

[3] Application Laboratory, Japan Agency for Marine-Earth Science and Technology, Yokohama, Japan.

Corresponding author: Takuro Matsuta (matsuta@ees.hokudai.ac.jp)

# ABSTRACT

Recent studies have reported that an increase in the bottom drag coefficient can enhance the volume transport of the Antarctic Circumpolar Current. Several mechanisms have been proposed to explain this frictional control, including the regulation of the geostrophic velocity by baroclinic instability and the influence of the form stress associated with standing meanders and wind-driven gyres. In this study, the role of momentum transport associated with Rossby wave radiations from disturbances is investigated as a potential frictional control mechanism. To highlight roles of the Rossby wave radiation, numerical experiments are conducted using barotropic reentrant channel models with topographic obstacles. In the high-drag regime, the circumpolar component is wind-driven, and the imbalance between the westerlies and topographic form stress sustains a net eastward transport. In contrast, in the low-drag regime, the eddy-driven westward circumpolar current is formed. In this case, the eastward flow at the center of the double gyre becomes unstable to barotropic instability. Analyses of the wave activity flux and momentum budget indicate that the Rossby wave transports westward momentum both northward and southward from the unstable region, which is responsible for the westward circumpolar current formation and maintenance. Although the direct application of the barotropic channel model to oceans requires caution, our findings imply that Rossby wave radiations from jets may play a role in the frictional control of the Antarctic Circumpolar Current.

## 1. Introduction

The Antarctic Circumpolar Current (ACC) is a wind-driven zonal jet around Antarctica, with a modest contribution from remote diapycnal mixing. Because the ACC directly links the three major ocean basins through the Drake Passage, understanding factors that determine its transport is essential for understanding the nature of the global ocean circulation (Talley 2013). Therefore, theories of wind-driven reentrant channel models have been proposed and investigated to explain dynamics of the ACC (e.g., Munk and Palmén 1951; Gill 1968; Johnson and Hill 1975; Webb 1993; Ishida 1994; Krupitsky and Cane 1994; Wang and Huang 1995; Uchimoto and Kubokawa 2005; Ward and Hogg 2011; Thompson and Sallée 2012; Abernathey and Cessi 2014; Constantinou and Young 2017; Bai et al. 2021; Matsuta and Mitsudera 2024; Zhang et al. 2024).

Recent studies have reported that bottom drag regulates circumpolar transport (Hogg and Blundell 2006; Uchimoto and Kubokawa 2005; Nadeau and Straub 2012; Nadeau and Ferrari 2015; Marshall et al. 2017; Constantinou 2018; Stewart and Hogg 2017; Matsuta and Mitsudera 2024). In both stratified and barotropic configurations, circumpolar transport increases as the linear bottom drag coefficient increases (Hogg and Blundell 2006; Nadeau and Straub 2012; Nadeau and Ferrari 2015; Marshall et al. 2017; Constantinou 2018; Matsuta and Mitsudera 2024). Several studies have explained these counterintuitive results based on eddy-mean flow interactions. According to Marshall et al. (2017), in a stratified channel, larger bottom drag requires a steeper isopycnal slope to sustain a sufficient eddy growth rate for the equilibration of meridional overturning. Consequently, circumpolar transport referenced to the bottom flow increases as the drag coefficient increases. In a barotropic configuration, if geostrophic contours are circumpolar and open, the increased bottom drag suppresses the barotropic instability that supports the phase difference between pressure fields and topography, thereby reducing the form stress (Constantinou 2018). In addition to the eddy-mean flow interactions, some studies suggested that standing Rossby waves (Stewart et al. 2023), a higher mode resonance (Uchimoto and Kubokawa 2005), and gyre circulations (Nadeau and Ferrari 2015; Matsuta and Mitsudera 2024) also contribute to the counterintuitive relationship.

An important aspect missing from the frictional control mechanisms is the representation of Rossby wave radiations from unstable jets. In western boundary current extensions, temporal fluctuations forced by barotropic instability or external forcing radiate barotropic Rossby waves and influence the formation and maintenance of recirculation gyres, even under stratified configurations (e.g., Provost and Verron 1987; Waterman and Jayne 2011; Mizuta 2012; Waterman and Jayne 2012; Waterman and Hoskins 2013; Mizuta 2018a,b). In addition, Provost and Verron (1987) demonstrated that wind-driven eastward flows indicate a transition from a steady state to an unstable state by reducing the linear bottom drag coefficient. These results suggest that the barotropic instability and associated Rossby wave radiation may contribute to the frictional control.

To test this hypothesis, a frictional control mechanism of circumpolar transport is investigated using barotropic reentrant channel models. Although the barotropic configuration is highly idealized, recent studies have suggested that even in stratified models, barotropic processes play an important role in adjustment processes of the ACC (Youngs et al. 2017; Jouanno and Capet 2020; Kong and Jansen 2021; Bai et al. 2021; Zhang et al. 2023, 2024). For example, Zhang et al. (2023) showed that changes in the barotropic flow modify the curvature of the standing meander. Another study (Youngs et al. 2017) pointed out that once the standing meander becomes sufficiently developed, the barotropic instability contributes to adjustment processes as well as baroclinic instability. However, these studies primarily focused on changes in westerly, and roles of barotropic instability in the adjustment processes in response to friction changes have not yet been explicitly discussed. Therefore, we expect that isolating the fundamental processes involving barotropic instability using a barotropic configuration provides a foundation for understanding frictional control in more complex stratified models. The key difference from previous studies of barotropic reentrant channel models (Uchimoto and Kubokawa 2005; Constantinou and Young 2017; Constantinou 2018; Constantinou and Hogg 2019) lies in the bottom topography. Earlier studies employed trains of low topographies and examined cases in which a standing Rossby wave resonated with the underlying topography or interacted with transient eddies. In contrast, this study employs a sufficiently high and isolated topography to locally block geostrophic contours. In this case, the mean flow behaves as wind-driven gyre circulations downstream of the topography (Patmore et al. 2019; Matsuta and Mitsudera 2024), similar to western boundary current extensions or the eddy hotspots in the ACC (Thompson and Naveira Garabato 2014; Matsuta and Masumoto 2023).

The rest of this paper is organized as follows: Model configurations are presented in Section 2. Section 3 presents the analysis of the dependencies of horizontal circulations and momentum balance on the bottom drag. In addition, the analysis of the wave activity flux is conducted and discussed to link eddy forcing with the Rossby wave radiation. Section 4 discusses the spinup processes to elucidate the mechanism responsible for the frictional control. In Section 5, we discuss the robustness of our proposed mechanism and its differences from other frictional control mechanisms. Section 6 summarizes the findings of the study and concludes the paper.

## 2. Model and Theory

### *2.1 Theoretical background*

#### 2.1.1 Momentum balance

In this study, an arbitrary variable $\ell$ is decomposed into time-averaged, *i.e.*, mean component $\overline{\ell}$ and transient eddy component $\ell'$ as a deviation from the mean component. With this separation, the mean momentum balance for the hydrostatic barotropic reentrant flow is governed by

$$\underbrace{\rho_0 \frac{\partial h\overline{\mathbf{u}}}{\partial t}}_{\text{TEND}} = \underbrace{-h\boldsymbol{\nabla}\overline{p}}_{\text{PGF}} + \underbrace{(-\rho_0 \boldsymbol{\nabla} \cdot h\overline{\mathbf{u}}\,\overline{\mathbf{u}})}_{\text{MEANADV}} + \underbrace{\left(-\rho_0 \boldsymbol{\nabla} \cdot h\overline{\mathbf{u}'\mathbf{u}'}\right)}_{\text{CONV_REY}} + \underbrace{\rho_0\{\mu_h h \nabla^2 \overline{\mathbf{u}} - r\overline{\mathbf{u}}\}}_{\text{DISP}} + \underbrace{(-\rho_0 f \mathbf{k} \times h\overline{\mathbf{u}})}_{\text{CORI}} + \underbrace{\boldsymbol{\tau}_{wind}}_{\text{WIND}} = 0, \quad (1)$$

where $\rho_0$ is the reference density; $h$ is the ocean thickness; $\mathbf{u} = (u, v)$ is the horizontal velocity, with $u, v$ being the x-direction (zonal) and y-direction (meridional), respectively; $p$ is the pressure, $\boldsymbol{\nabla} = (\partial_x, \partial_y)$ is the gradient operator, with $\partial_\#$ being the partial derivative of subscript #; $f = f_0 + \beta y$ is the Coriolis parameter with the $\beta$ coefficient; $\mu_h$ is the horizontal viscosity per density; $\mathbf{k}$ is the unit vector pointing upward; and the steady wind stress $\boldsymbol{\tau}_{wind} = (\tau_x, \tau_y)$. The implicit linear free-surface condition is applied at the ocean surface. The lefthand side corresponds to the tendency (TEND). On the righthand side, from left to right, the terms correspond to the pressure gradient force (PGF), advection of the mean flow (MEANADV), the convergence of the Reynolds stress (CONV_REY), momentum dissipation due to the bottom drag and horizontal viscosity (DISP), the Coriolis term (CORI), and the wind stress (WIND). When the momentum balance is integrated in the zonal direction, the PGF corresponds to the topographic form stress (TFS), while the Coriolis term vanishes in the steady state.

#### 2.1.2 Potential vorticity equation and barotropic instability

This study examines how the bottom drag alters the horizontal distribution of the single-layer quasi-geostrophic potential vorticity (PV).

$$q(x,y) = \zeta + \beta y + \frac{f_0}{h}\eta_b, \tag{2}$$

where $\zeta = \nabla \times \mathbf{u}$ is the relative vorticity and $\eta_b$ is the topographic height. Here, we neglect the contribution of the sea surface elevation $f_0\eta/h$ because the deformation radius is sufficiently large. The last two terms of the PV, $\beta y + (f_0/h)\eta_b$ correspond to the geostrophic contour of the quasi-geostrophic framework. The quasi-geostrophic approximation is inapplicable in regions where the topographic height variation is not small; however, in this study, the focus is on the downstream region of the topography where $\eta_b/h$ is sufficiently small. Thus, the quasi-geostrophic framework is adopted. According to certain studies (e.g., Pedlosky 1987; Vallis 2017), a necessary condition of barotropic instability is that the meridional gradient of the PV changes sign somewhere within the domain.

#### 2.1.3 Wave activity flux

For small-amplitude transient eddies on a slowly varying barotropic time-mean flow, the growth and decay of eddies are approximately regulated by the wave activity equation:

$$\frac{\partial M}{\partial t} + \nabla \cdot (\bar{\mathbf{u}}M) + \nabla \cdot \mathbf{W} = D, \tag{3}$$

where $M = 0.5\left(\overline{q'^2}/|\nabla\bar{q}|\right)$ is the pseudo-momentum or wave activity density and $D$ is the wave activity flux dissipation (Plumb 1986). The radiative wave activity flux $\mathbf{W}$ is defined by

$$\mathbf{W} = \frac{\gamma}{\sqrt{\bar{u}^2 + \bar{v}^2}} \begin{pmatrix} \frac{\bar{u}}{2}\left(\overline{v'^2} - \overline{u'^2}\right) - \bar{v}\overline{u'v'} \\ \frac{\bar{v}}{2}\left(-\overline{v'^2} + \overline{u'^2}\right) - \bar{u}\overline{u'v'} \end{pmatrix}, \tag{4}$$

where $\gamma = 1$ if the mean flow is pseudo-eastward such that the PV increases to the right of the mean flow; the opposite ($\gamma = -1$) happens if it is pseudo-westward. Because the PV increases northward in our case, as discussed in Section 3.1, we assume $\gamma = 1$ when the zonal component of velocity is positive; the opposite happens if zonal component is negative. An advantage of the wave activity flux concept is that, in the quasi-plane-wave limit, the radiative wave activity flux becomes parallel to the group velocity ($\mathbf{c}_g$) of the Rossby wave relative to the mean flow, $\mathbf{W} = (\mathbf{c}_g - \bar{\mathbf{u}})M$. Hereafter, we call $\mathbf{W}$ the "wave activity flux" for simplicity.

The divergence and convergence of the wave activity flux describe the eddy feedback on the mean flow. According to Plumb (1986), the wave activity flux corresponds to the effective eddy flux of pseudo-westward momentum; hence, the convergence of the wave activity flux works as the westward forcing on the mean flow, whereas the opposite is true for the divergence. Therefore, the analysis of the wave activity flux and its convergence demonstrates the Rossby wave propagation and associated momentum transport. One may argue that applying the wave activity concept is unfavorable because the jet consists of nonlinear eddies. However, transient Rossby waves emitted from unstable areas may be regarded as linear waves. The wave activity flux concept successfully describes eddy-mean flow interactions in unstable ocean jets (e.g., Thompson and Naveira Garabato 2014; Chapman et al. 2015; Foppert 2019; Matsuta and Masumoto 2021).

*2.2 Model description*

In this study, the idealized wind-driven barotropic reentrant flow on the beta-plane channel was solved using the general circulation model of the Massachusetts Institute of Technology (MITgcm, Marshall et al. (1997)). Our configurations were similar to those of Matsuta and Mitsudera (2024). The domain was $L_x = 6000$ km long (zonal, x-direction) and $H_0 = 4000$ m deep (vertical, z-direction). The meridional (y-direction) extent, $L_y$ is defined in the next paragraph. The free-slip boundary condition was imposed at the northern and southern boundaries. A Cartesian grid with a horizontal resolution of $5 \times 5$ km was used. The Coriolis parameter is defined by $f = f_0 + \beta y$, with $f_0 = -10^{-4}$ s$^{-1}$ and $\beta = 10^{-11}$ m$^{-1}$s$^{-1}$. The reference density $\rho_0$ was $1035$ kg m$^{-3}$. The subgrid horizontal eddy viscosity, $\mu_h$ was $12$ m$^2$ s$^{-1}$.

A central point of this work was to investigate the dependencies of circumpolar transport on the bottom drag. We changed the bottom drag linear coefficient, $r = \{10^{-4}, 10^{-3}\}$ [m s$^{-1}$]. The models with $r = 10^{-4}$ m s$^{-1}$ and $r = 10^{-3}$ m s$^{-1}$ are called "LOW-DRAG" and "HIGH-DRAG," respectively. In addition, a strong-viscosity experiment denoted as "VISC" was conducted. In this configuration, the horizontal viscosity is $\mu_h = 2000$ m s$^{-1}$, and the other configurations are the same as those of LOW-DRAG. Because the barotropic instability is suppressed in this experiment, the comparison between the LOW-DRAG and VISC experiments highlights roles of barotropic instability. Furthermore, a "VISC+EDDY" experiment was conducted. In this experiment, the eddy forcing diagnosed in the LOW-DRAG experiment was added to the wind forcing as external forcing, i.e., $\boldsymbol{\tau}_{wind}$ is replaced by $\boldsymbol{\tau}_{wind} + \left(-\rho_0 \nabla \cdot h\overline{\mathbf{u}'\mathbf{u}'}\right)_{\mathrm{diagnosed}}$, and the other conditions were the same as those of VISC. The added eddy forcing did not alter the total momentum gain. Each configuration is listed in Table 1.

Because Constantinou (2018) suggested that the geometries of geostrophic contours possibly alter the nonlinear barotropic dynamics, the LOW-DRAG and HIGH-DRAG models were investigated using three types of topographies. The first type of topography (called Type A, hereafter) was the Gaussian ridge centered at $x_0 = 1000$ km:

$$\eta_b(x, y) = \eta_0 \exp\left[-\frac{(x - x_0)^2}{\sigma_x^2}\right], \tag{5}$$

where $\eta_0 = 1500$ m was the maximum height of topography and $\sigma_x = 300$ km was the zonal width of the topography (Figure 1a). In this case, the meridional extent of the domain was $L_y = 2000$ km. Because the topography was connected to the northern and southern walls, all the geostrophic contours were blocked. Here, the geostrophic contour was calculated by $f/h$. The zonal wind forcing for Type A is expressed as follows:

$$\tau_w(y) = \begin{cases} \tau_0 \sin^2\left(\pi \dfrac{y - y_3}{y_4 - y_3}\right), & y_3 \le y \le y_4 \\ 0, & \text{otherwise} \end{cases}, \tag{6}$$

where $\tau_0 = 0.4$ N m$^{-2}$, $y_3 = 500$ km and $y_4 = 1500$ km (Figure 1(b)).

The second type of topography (called Type B, hereafter) was an isolated Gaussian ridge centered at $x_0 = 1000$ km:

$$\eta_b(x, y) = \begin{cases} \eta_0 \exp\left[-\dfrac{(x - x_0)^2}{\sigma_x^2} - \dfrac{(y - y_0)^2}{\sigma_y^2}\right], & 0 \le y \le y_0 \\ \eta_0 \exp\left[-\dfrac{(x - x_0)^2}{\sigma_x^2}\right], & y_0 \le y \le y_1 \\ \eta_0 \exp\left[-\dfrac{(x - x_0)^2}{\sigma_x^2} - \dfrac{(y - y_1)^2}{\sigma_y^2}\right], & y_1 \le y \le L_y \end{cases}, \tag{7}$$

where $L_y = 3000$ km, $\eta_0 = 1500$ m, $y_0 = 500$ km, $y_1 = 2500$ km, $\sigma_x = 300$ km was the zonal width of the topography, and $\sigma_y = 150$ km was the width of the meridional slopes connected to the northern and southern tips of the topography (Figure 1(c)). As shown in Figure 1(c), some geostrophic contours were circumpolar, whereas others were closed. The zonal wind forcing for Type B is the same as (6) but $y_3 = 1000$ km and $y_4 = 2000$ km (Figure 1(d)). It is noted that the wind stress was shifted meridionally relative to Type A to align it with the topographic center, but its width and amplitude were unchanged, so the total momentum input remained the same.

The shape of the third type of topography (called Type C, hereafter) was the same as that of Type A, but the model domain was extended from $L_y = 2000$ km to $L_y = 5000$ km northward (Figure 1(e)). Consequently, some geostrophic contours became circumpolar. The wind stress was the same as that in the Type B case (Figure 1(f)).

The model reached a quasi-equilibrium state after the fifth year, and the transport characteristics did not change significantly thereafter. Therefore, the period from the 10th year to the 12th year was used for the analysis. Unless otherwise stated, the time mean over this period is defined as the mean field, and the deviation from it is defined as the eddy component.

Table 1 List of model configurations showing model abbreviations, values of the drag coefficient and viscosity, and the existence of diagnosed eddy forcing.

| Model | $r$ [m s$^{-1}$] | $\nu$ [m$^2$ s$^{-1}$] | Diagnosed eddy forcing |
|---|---|---|---|
| LOW-DRAG | $10^{-4}$ | 12 | |
| HIGH-DRAG | $10^{-3}$ | 12 | |
| VISC | $10^{-4}$ | 2000 | |
| VISC+EDDY | $10^{-4}$ | 2000 | ✓ |

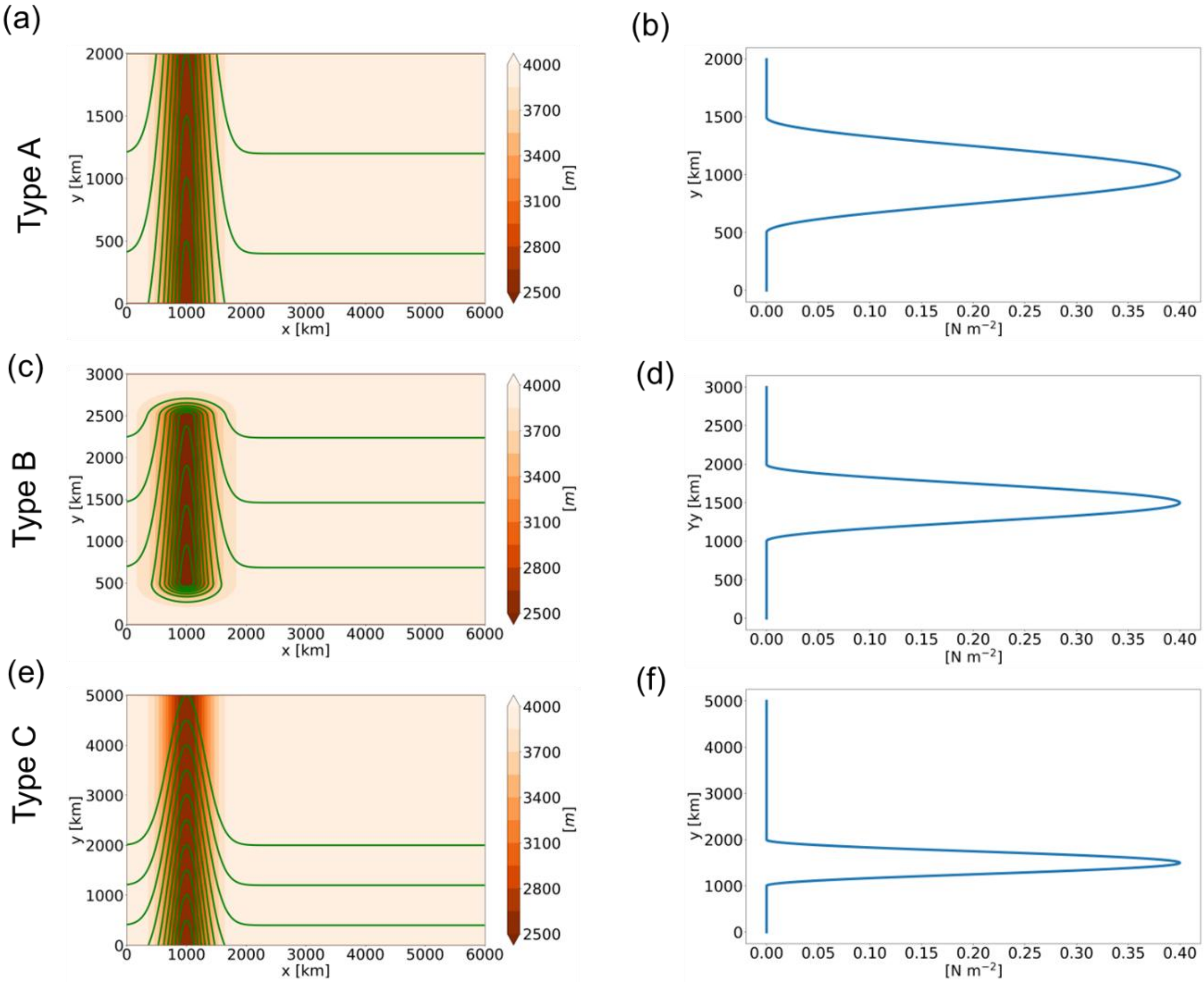

Figure 1. (a) Color shades representing a horizontal distribution of the ocean depth of Type A. Green contours denote the geostrophic contours with a contour interval of $2.0 \times 10^{-9}$ m$^{-1}$ s$^{-1}$. (b) Meridional profile of wind stress for Type A. (c) and (d) are the same as (a) and (b) but for Type B, respectively. (e) and (f) are the same as (a) and (b) but for Type C, respectively.

## 3. Result

*3.1 Horizontal structure of circumpolar currents*

First, horizontal circulations are diagnosed. In Figure 2, the PV (Eq. (2)) and the barotropic streamfunction are illustrated. Here, the streamfunction is determined as

$$\psi(x,y) = \int_0^y hu(x,y)dy. \tag{8}$$

The unit of $\psi$ is $\mathrm{Sv} = 10^6\ \mathrm{m^3\ s^{-1}}$ in this case. If $\psi$ at the northern boundary is positive, the net transport is eastward, and if it is negative, it is westward. Figure 2(a), (b) and (c) illustrate a critical difference between the LOW-DRAG and HIGH-DRAG configurations of Type A: the LOW-DRAG case exhibits a westward circumpolar current, whereas the HIGH-DRAG case exhibits an eastward circumpolar transport. In LOW-DRAG, the circumpolar current goes westward south of $y = 500$ km within the region of $2000\ \mathrm{km} \leq x \leq 6000$ km and then turns northward along the equivalent western boundary centered at $x = 1000$ km. This westward component traverses the geostrophic contours at the northern wall, resulting in a net westward transport. Conversely, in HIGH-DRAG, the weak eastward circumpolar component indicated by the 0 Sv contour goes over the topography via the northern boundary. Furthermore, gyre circulations in LOW-DRAG are stronger than those in HIGH-DRAG. For example, while the maximum transport of the southern gyre is −160 Sv in HIGH-DRAG, it exceeds −600 Sv in LOW-DRAG. Moreover, the anticyclonic circulation is formed over the topography in LOW-DRAG. This is because the temporal fluctuations around the sea mount in the beta plane generate mean anticyclonic vortices due to the PV conservation (e.g., Whitehead 1975; Salmon 1998).

Simulation results of Type B and C are the same as those of Type A. Figure 2(d) and Figure 2(f) show that the westward circumpolar currents are formed northward of the northern gyre (indicated streamfunctions between −600 Sv and −240 Sv) and southward of the southern gyre (indicated by the −120 Sv streamfunction contours) in the LOW-DRAG configuration of Type B. In contrast, the HIGH-DRAG configuration shows an eastward circumpolar current indicated by streamfunction contours between 0 Sv and 40 Sv (Figure 2(e)). The situations are common in the Type C configuration (Figure 2(g-i)). Because the situations are not dependent on the topography, the study focuses on the Type B models as representative cases, hereafter.

To quantitatively represent the relationship between the circumpolar transport and drag coefficient, a plot of the circumpolar transport as a function of the drag coefficient is illustrated in Figure 3(a). Here, we perform a suite of sensitivity experiments with varying drag coefficients in addition to the LOW-DRAG and HIGH-DRAG cases. The transport is eastward when the drag coefficient is larger than $\sim 10^{-3}$ m $\mathrm{s}^{-1}$, otherwise it shifts westward at around $2.8 \times 10^{-4}$ m $\mathrm{s}^{-1}$. The plot appears to show a jump due to the use of a symmetric logarithmic scale, but the transition is continuous. When the drag coefficient is set near transition values, the transport oscillates around approximately $\pm 10$ Sv, and the sign of its annual mean sometimes varies (not shown). This behavior indicates that the transition occurs continuously rather than through a noncontinuous bifurcation. Figure 3(b) demonstrates that reversing the circumpolar transport is accompanied by a significant increase in the eddy kinetic energy (EKE), $0.5\left(\overline{u'^2} + \overline{v'^2}\right)$. When the transport is eastward, the EKE is approximately zero, but it increases as the westward flow strengthens.

A comparison of the meridional PV distribution further emphasizes roles of eddies. The meridional profile of the PV along $x = 4000$ km is used as an example. In LOW-DRAG (Figure 4 (a)), the PV is partially homogenized at approximately $y = 1000$ km and $y = 2000$ km, which characterizes barotropic instability. Reflecting the PV staircase structure, the eastward flow is sharp at the center, whereas the westward flows have a parabolic shape and are relatively broad. Conversely, in the HIGH-DRAG case (Figure 4 (b)), the PV is not homogenized. Because the meridional PV gradient is always positive, the high-drag regime is stable to barotropic instability. The HIGH-DRAG model is in the steady state as shown in Figure 3(b). These differences suggest that barotropic instability is responsible for the maintenance of the westward circumpolar current.

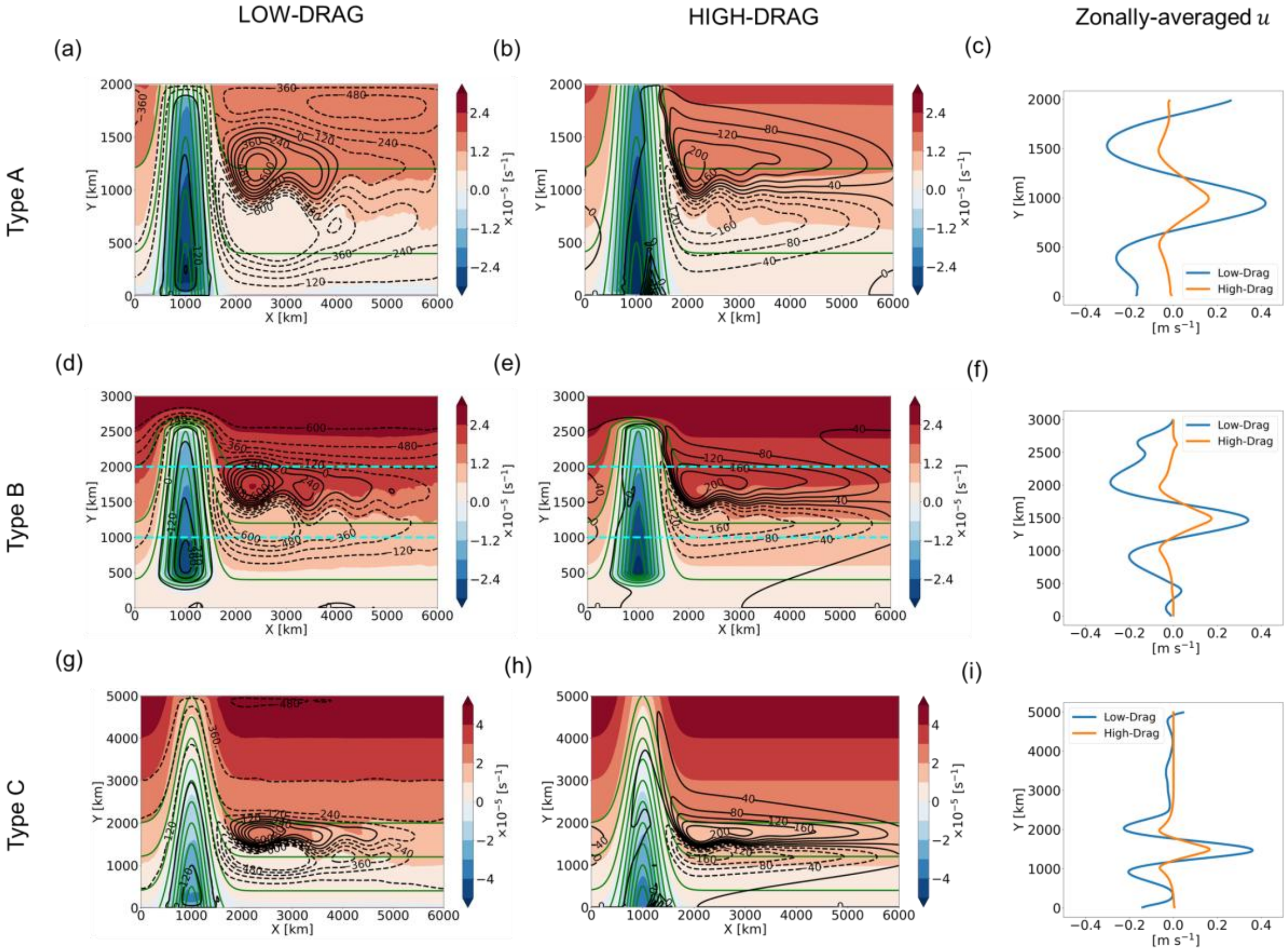

Figure 2. Horizontal distribution of PV and streamfunctions in (a) LOW-DRAG and (b) HIGH-DRAG for Type A. Color shades indicate the horizontal distribution of PV, and black contours indicate the streamfunctions. The contours larger than 600 Sv are not shown. Green contours denote the geostrophic contours with a contour interval of $2.0 \times 10^{-9}$ m$^{-1}$ s$^{-1}$. (c) Meridional profile of zonally-averaged zonal velocity for LOW-DRAG (blue curve) and HIGH-DRAG (orange curve). The middle column are the same as the upper column but for Type B. Cyan dashed lines are defined in Section 3.2. The lower column are the same as the upper column but for Type C.

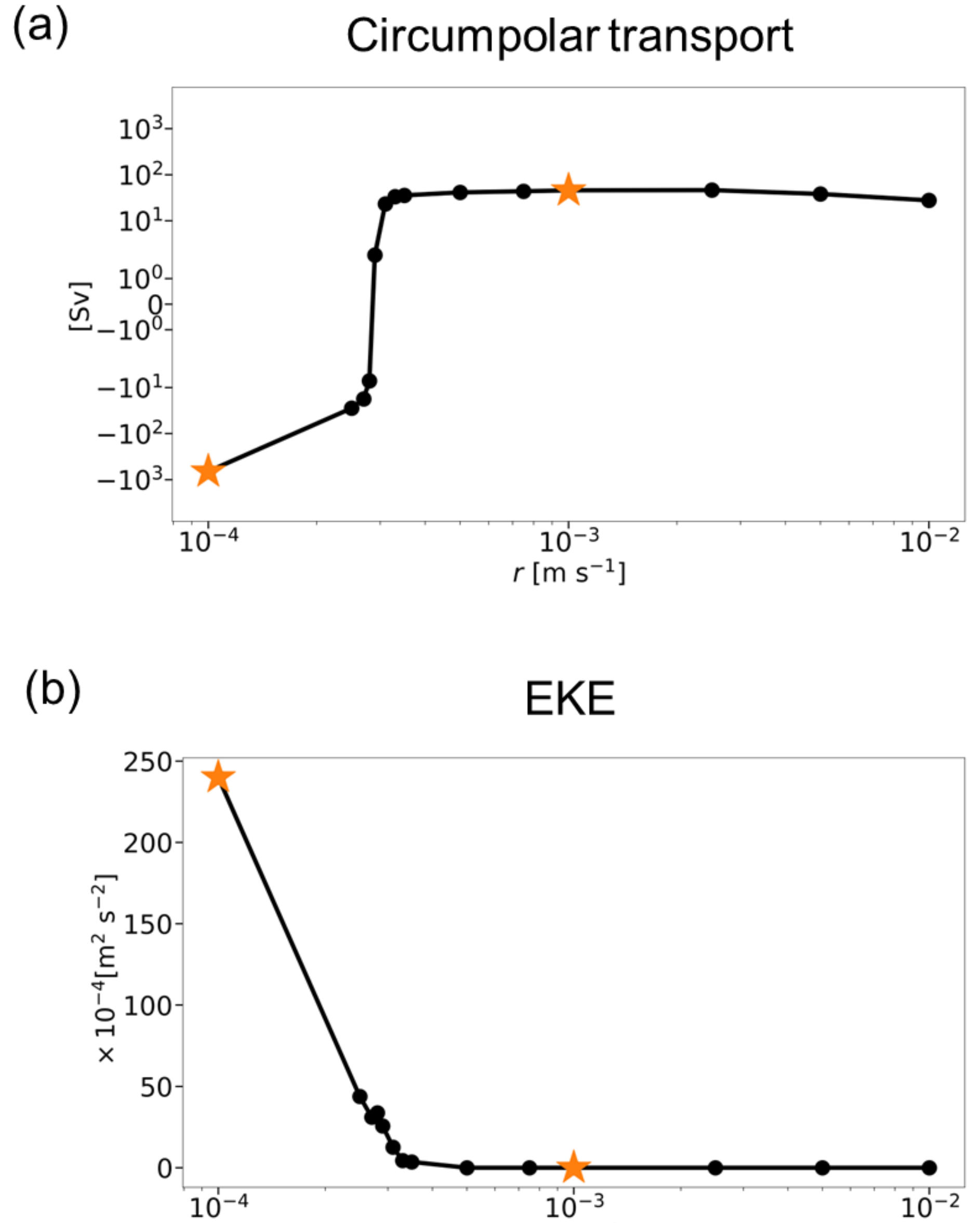

Figure 3 (a) Circumpolar transport and (b) domain-averaged EKE as functions of drag coefficients. LOW-DRAG and High-DRAG cases are marked by yellow stars.

(a) LOW-DRAG

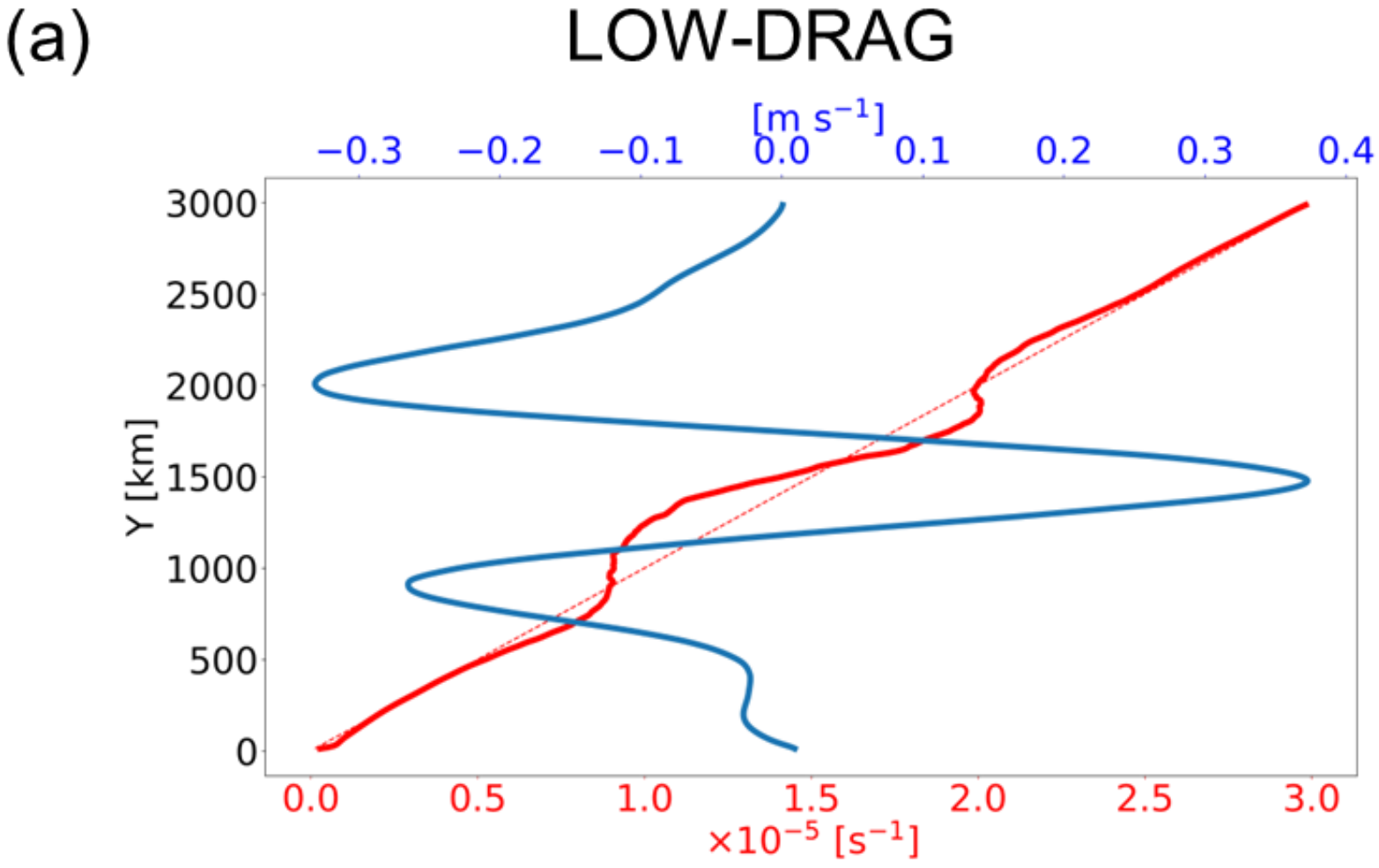

(b) HIGH-DRAG

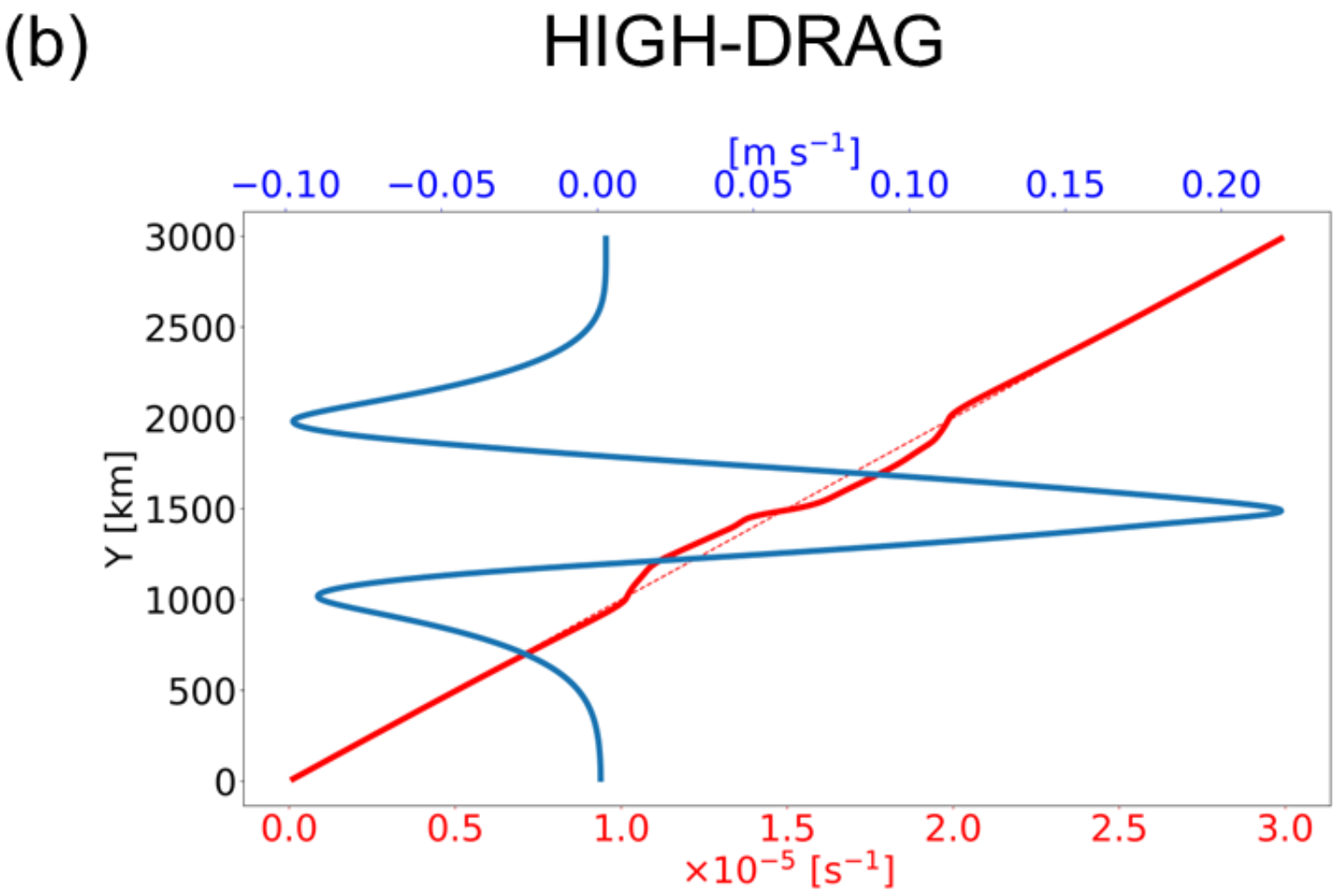

Figure 4 (a) Meridional profile of PV (red line) and the zonal velocity (blue line) along $x = 4000$ km in LOW-DRAG. The red dashed line indicates $\beta y$. (b) Same as (a) but for HIGH-DRAG.

*3.2 Momentum budget*

Figure 5 shows the meridional profile of the zonally-averaged zonal momentum budget for LOW-DRAG and HIGH-DRAG. The tendency term is much smaller than the other terms, and the Coriolis term is theoretically zero in a steady state; therefore, these terms are not shown. The momentum balance of the cases shows two notable differences. The first is the dissipation term (DISP): In LOW-DRAG, corresponding to a smaller drag coefficient, the DISP term in the wind-forced region is reduced to approximately one-tenth, compared to that in HIGH-DRAG. It is noted that although the magnitude of friction itself is large in HIGH-DRAG, the DISP term integrated over the domain accounts for only about 6.6% of the wind stress and is therefore consistent with the momentum balance reported in previous studies (Masich et al. 2015; Constantinou and Hogg 2019; Patmore et al. 2019). Therefore, the sensitivity of transport on friction is small for $r \geq 4 \times 10^{-4}$ m s$^{-1}$ (Figure 3). The second difference is the presence of the Reynolds stress term (CONV_REY). As shown in Section 3.1, the HIGH-DRAG model reaches a steady state, making the Reynolds stress term equals to zero. In contrast, in LOW-DRAG, the Reynolds stress converges between $y = 1000$ km and $y = 1500$ km while diverging adjacent to this region. This profile behavior characterizes momentum redistribution by barotropic Rossby wave radiations from disturbances (Vallis 2017). The pathways of the Rossby wave propagation will be discussed in the next section and Appendix.

The mean advection (MEANADV) also contributes to the redistribution of momentum, a feature that is common to the HIGH-DRAG case as well. This momentum redistribution is associated with the western boundary current. The northwestward flow associated with the southern gyre transports westward momentum northward, whereas the southeastward flow associated with the northern gyre transports eastward momentum southward. Consequently, the mean advection works as the eastward forcing south of $y = 1500$ km and westward forcing north of $y = 1500$ km.

To elucidate roles of the momentum redistributions, the momentum budget is conducted over the region north of $y = 2000$ km (Region N), between $y = 1000$ km and $y = 2000$ km (Region C), as well as south of $y = 1000$ km (Region S) for the LOW-DRAG case. The boundaries of these areas are indicated by cyan dashed lines in Figure 2(d) & Figure 2(e). Table 2 demonstrates that the momentum redistributions substantially alter the meridional profiles of the topographic form stress (TFS) and DISP. In Region C, CONV_REY and MEANADV work as eastward forcing, indicating that these terms remove the westward momentum from this area.

Because the bottom drag coefficient is small, the eastward forcings by the advection terms are mainly balanced by the increased TFS. The westward momentum removed from Regions C and S by MEANADV converges in Region N. The westward forcing, with a magnitude of $-12 \times 10^{10}$ N is balanced by a positive TFS with a value of $11 \times 10^{10}$ N, suggesting that the MEANADV mainly alters the meridional profile of the TFS and has little impacts on the momentum budget integrated over the domain.

By contrast, the westward momentum transported by the Reynolds stress convergences in Regions N and S: the associated westward forcing is $-8.8 \times 10^{10}$ N in Region N, whereas it is $-5.2 \times 10^{10}$ N in Region S. The converged westward momentum is roughly balanced by DISP, with values of $9.5 \times 10^{10}$ N and $4.0 \times 10^{10}$ N in Regions N and S, respectively. Unlike the MEANADV, the momentum transport by CONV_REY does not simply alter the meridional profiles of the TFS and DISP. In Region C, the eastward eddy forcing is balanced by the TFS, whereas the converged westward momentum is balanced by the DISP in Regions N and S. Because the eastward drag in Regions N and S remains even after integration over the entire domain, it is not acting solely on the return flow of the gyres, but rather on the net westward circumpolar flow. In other words, the balances between CONV_REY and DISP in Regions N and S sustain the net westward circumpolar transport. Although the eddy forcing integrates to zero over the entire domain, it catalyzes the net changes of the TFS and DISP in the total momentum balance, ultimately shifting the direction of the circumpolar current.

For comparison, the momentum budget of the HIGH-DRAG case (Table 3) is discussed. As in the case of LOW-DRAG, the MEANADV works as the eastward forcing in Region C, while it works as the westward forcing in Region N. In this configuration, the converged westward momentum by MEANADV is balanced by DISP in Region N. However, as shown in Figure 2(d), the westward circumpolar component is not formed in Region N, and this friction acts on the return flow of the gyre. According to Matsuta and Mitsudera (2024), in this regime the MEANADV acts to intensify the magnitude of the TFS; however, its influence is insufficient to reverse the flow direction. This suggests that MEANADV does not serve as a forcing mechanism for westward circumpolar transport but merely modifies the mean pressure distribution. This point will be discussed further in Section 3.4.

Table 2 Each term of the momentum budget integrated over Regions N, C, and S, and the whole domain for the LOW-DRAG case. TFS, MEANADV, CONV_REY, DISP, and WIND denote the topographic form stress, advection of mean flow, convergence of Reynolds stress, momentum dissipation, and wind stress, respectively. If imbalances exist, they are from the roundoff errors or small values of tendency and Coriolis force.

| $10^{10} \times$ [N] | TFS | MEANADV | CONV_REY | DISP | WIND |
|---|---|---|---|---|---|
| Region N | 11 | −12 | −8.8 | 9.5 | 0.0 |
| Region C | −140 | 7.4 | 14 | −2.8 | 119 |
| Region S | −3.7 | 4.8 | −5.2 | 4.0 | 0.0 |
| Total | −132 | $< 10^{-6}$ | $< 10^{-3}$ | 11 | 119 |

Table 3 Same as Table 2 but for the HIGH-DRAG case.

| $10^{10} \times$ [N] | TFS | MEANADV | CONV_REY | DISP | WIND |
|---|---|---|---|---|---|
| Region N | −2.9 | −2.8 | $< 10^{-6}$ | 5.8 | 0.0 |
| Region C | −101 | 3.1 | $< 10^{-6}$ | −22 | 119 |
| Region S | −8.2 | −0.23 | $< 10^{-6}$ | 8.5 | 0.0 |
| Total | −112 | $< 10^{-6}$ | $< 10^{-6}$ | −7.4 | 119 |

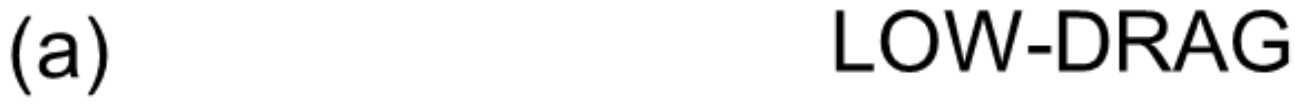

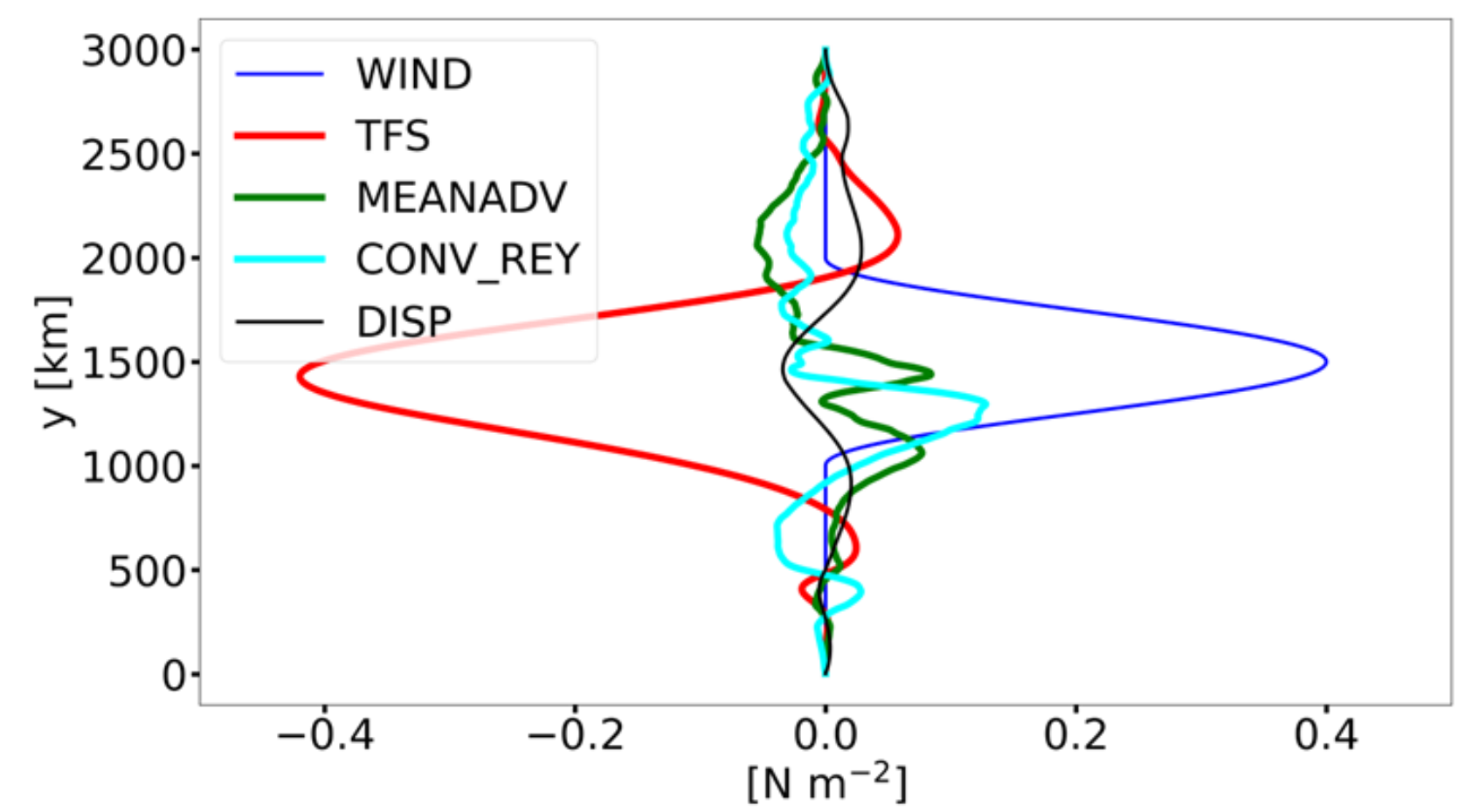

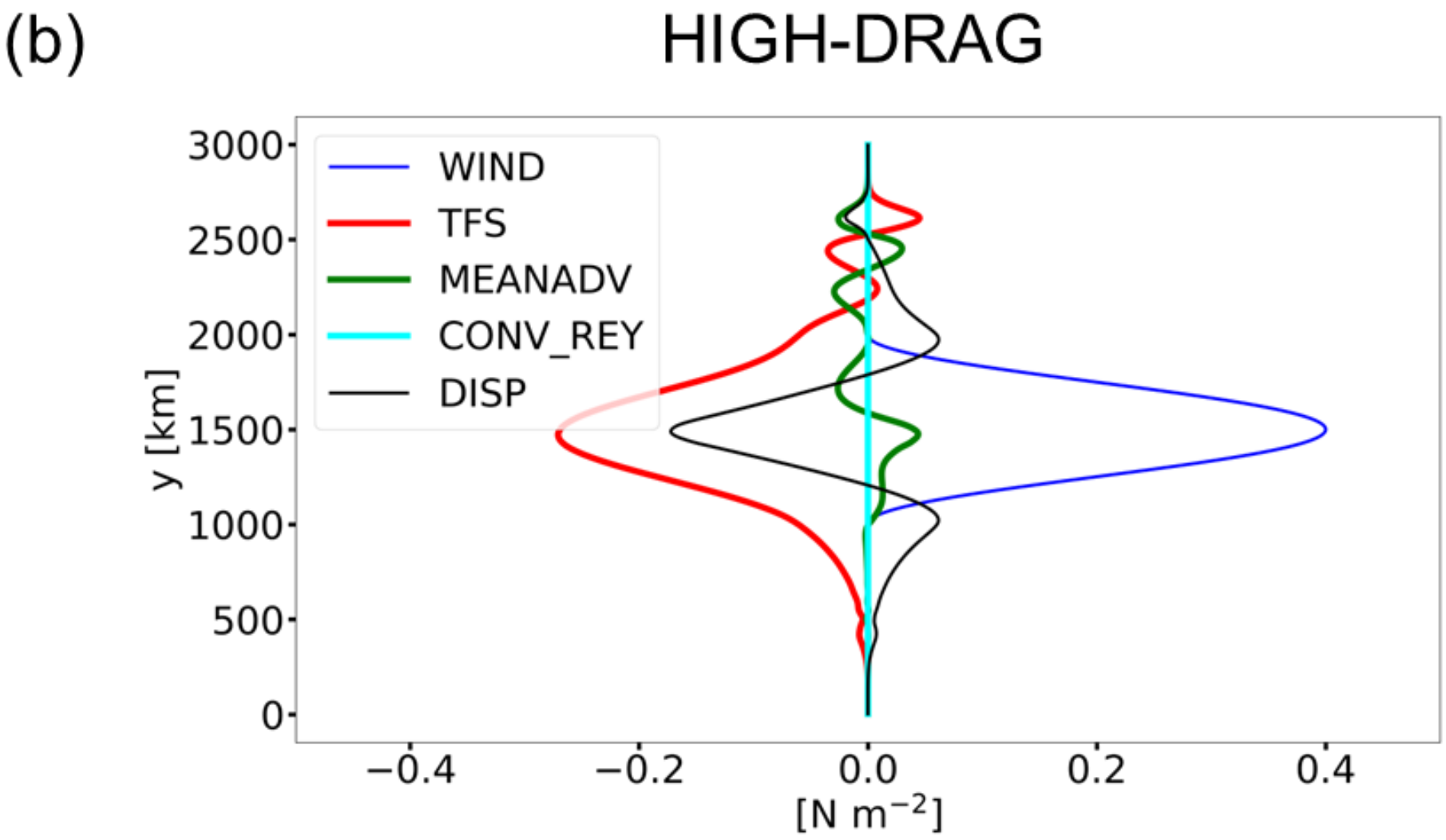

Figure 5. Meridional profiles of the zonally-averaged zonal momentum budget for (a) LOW-DRAG and (b) HIGH-DRAG. Blue, red, green, cyan, and black solid curves indicate the wind stress (WIND), topographic form stress (TFS), mean advection (MEANADV), convergence of the Reynold stress (CONV_REY), and momentum dissipation by friction and viscosity (DISP), respectively.

*3.3 Analysis of the wave activity flux*

The wave activity flux and EKE are analyzed to identify routes of westward momentum transport by the Rossby wave. The correspondence between the wave activity flux observed below and the propagation of Rossby waves was verified in Appendix. Figure 6(a) demonstrates that the EKE has large values downstream of $x = 3000$ km along $y = 1500$ km. The wave activity is radiated from this eddy-rich region. At the northern flank of the disturbances, the wave activity flux points northward, whereas the flux points southward at the southern flank. This distribution exhibits a structure that is characteristic of barotropic Rossby wave radiations caused by disturbances. (Vallis 2017). The wave activity is also radiated along the northern boundary of the northern gyre and southern boundary of the southern gyre (Provost and Verron 1987), although the amplitude is small. Consequently, the radiated wave activity converges into the outside of the wind-driven circulations indicated by the green boxes in Figure 6(b), which acts as a westward forcing on the mean flow. The locations where the wave activity flux converges correspond closely to those of the westward circumpolar current, suggesting that the westward pseudo-momentum associated with the Rossby wave radiations sustains the westward circumpolar transport.

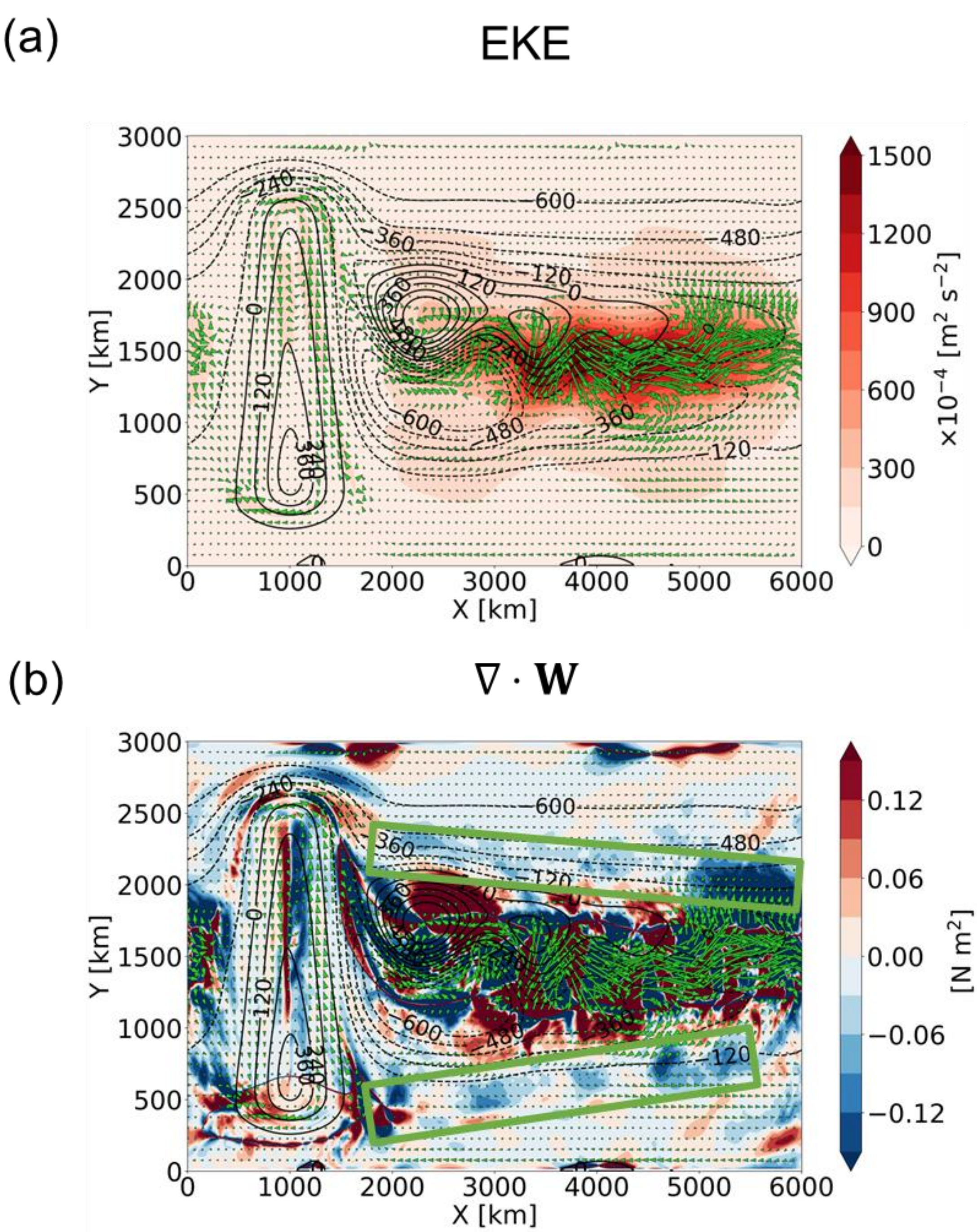

Figure 6 Color shades indicating the horizontal distribution of (a)EKE and the (b) divergence of the wave activity flux in LOW-DRAG. Black contours indicate the streamfunctions. Green arrows indicate the wave activity flux, **W**. Green rectangular boxes in (b) indicate the regions where the wave activity flux converges.

*3.4 Eddy forcing versus mean advection*

The VISC case was further investigated to confirm that the eddy forcing is essential and the mean advection is not responsible for the westward circumpolar current formation. The horizontal distribution of the streamfunction is shown in Figure 7(a), indicating two maxima in the northern gyres, which is similar to those of LOW-DRAG (Figure 2(c)). By contrast, the flow is in the steady state as in the HIGH-DRAG case. The absence of eddy activity is demonstrated in the two ways: there is no anticyclonic circulation over the topography, and, more importantly, the net circumpolar transport is eastward, with a magnitude of 28 Sv.

To quantitatively demonstrate the impact of the absence of eddy forcing on the flow, the momentum budget is discussed. Table 4 demonstrates that values of MEANADV and DISP are approximately equal to those in LOW-DRAG in Region C. The magnitude of the TFS decreases by $17 \times 10^{10}$ N, which can be attributed to the absence of the eastwards eddy forcing. Although the westward momentum converges in Region N, it is balanced by the TFS; hence, the MEANADV has little impact on the total momentum budget as in the case of LOW-DRAG. This result indicates that the MEANADV cannot form the westward circumpolar current.

To further emphasize the role of Reynolds stress, the VISC+EDDY case was investigated. As in the case of VISC, the flow reaches the steady state because the strong viscosity dumps the barotropic instability. However, the eddy forcing diagnosed from LOW-DRAG significantly alters the mean flow, even though it vanishes in the total momentum balance. As shown in Figure 7(b), the westward circumpolar current is formed north of $y = 2000$ km: the net transport is $-275$ Sv. In addition, the anticyclonic circulation over the topography is recovered, although the maximum value of 120 Sv is smaller than that in the LOW-DRAG case. The momentum balance (Table 5) also emphasizes the role of eddy forcing. Here, the diagnosed eddy forcing is included in WIND. In Region C, the magnitude of the TFS increases by the inclusion of the eastward eddy forcing. In contrast, in Regions N and S, the westward eddy forcing is mainly balanced by DISP. This momentum balance feature is similar to that in the LOW-DRAG case, indicating that eddy forcing behaves as an external forcing on the mean flow and alters it.

Based on the analyses conducted from Section 3.2 to this section, the mechanism maintaining the westward flow can be considered as follows: Disturbances associated with the barotropic instability generates Rossby waves in the latitude band between $y = 1000$ km and $y = 2000$ km, which transports westward momentum from the region of disturbances. The wave activity flux converges outside of the stirring region and works as the effective westward forcing for the mean flow, which sustains the westward circumpolar current. Although the mean advection redistributes the momentum in the meridional direction, this process alters the meridional profile of the bottom pressure field and has no substantial effects on the momentum balance integrated over the domain. The effective eddy forcing acting as an *external forcing* on the mean flow is essential for the westward circumpolar transport.

Table 4 Same as Table 2 but for the VISC case.

| $10^{10} \times$ [N] | TFS | MEANADV | CONV_REY | DISP | WIND | TEND |
|---|---|---|---|---|---|---|
| Region N | 6.0 | −7.5 | $< 10^{-6}$ | 1.5 | 0.0 | $< 10^{-6}$ |
| Region C | −123 | 7.0 | $< 10^{-6}$ | −3.0 | 119 | $< 10^{-6}$ |
| Region S | −1.5 | 0.47 | $< 10^{-6}$ | 1.0 | 0.0 | $< 10^{-6}$ |
| Total | −119 | $< 10^{-6}$ | $< 10^{-6}$ | −0.45 | 119 | $< 10^{-6}$ |

Table 5 Same as Table 2 but for the VISC+EDDY case. WIND includes the diagnosed eddy forcing in addition to the westerlies.

| $10^{10} \times$ [N] | TFS | MEANADV | CONV_REY | DISP | WIND | TEND |
|---|---|---|---|---|---|---|
| Region N | 11 | −7.6 | $< 10^{-6}$ | 5.3 | −8.8 | $< 10^{-6}$ |
| Region C | −131 | 5.5 | $< 10^{-6}$ | −7.1 | 133 | $< 10^{-6}$ |
| Region S | −3.0 | 2.0 | $< 10^{-6}$ | 6.2 | −5.2 | $< 10^{-6}$ |
| Total | −124 | $< 10^{-6}$ | $< 10^{-6}$ | 4.4 | 119 | $< 10^{-6}$ |

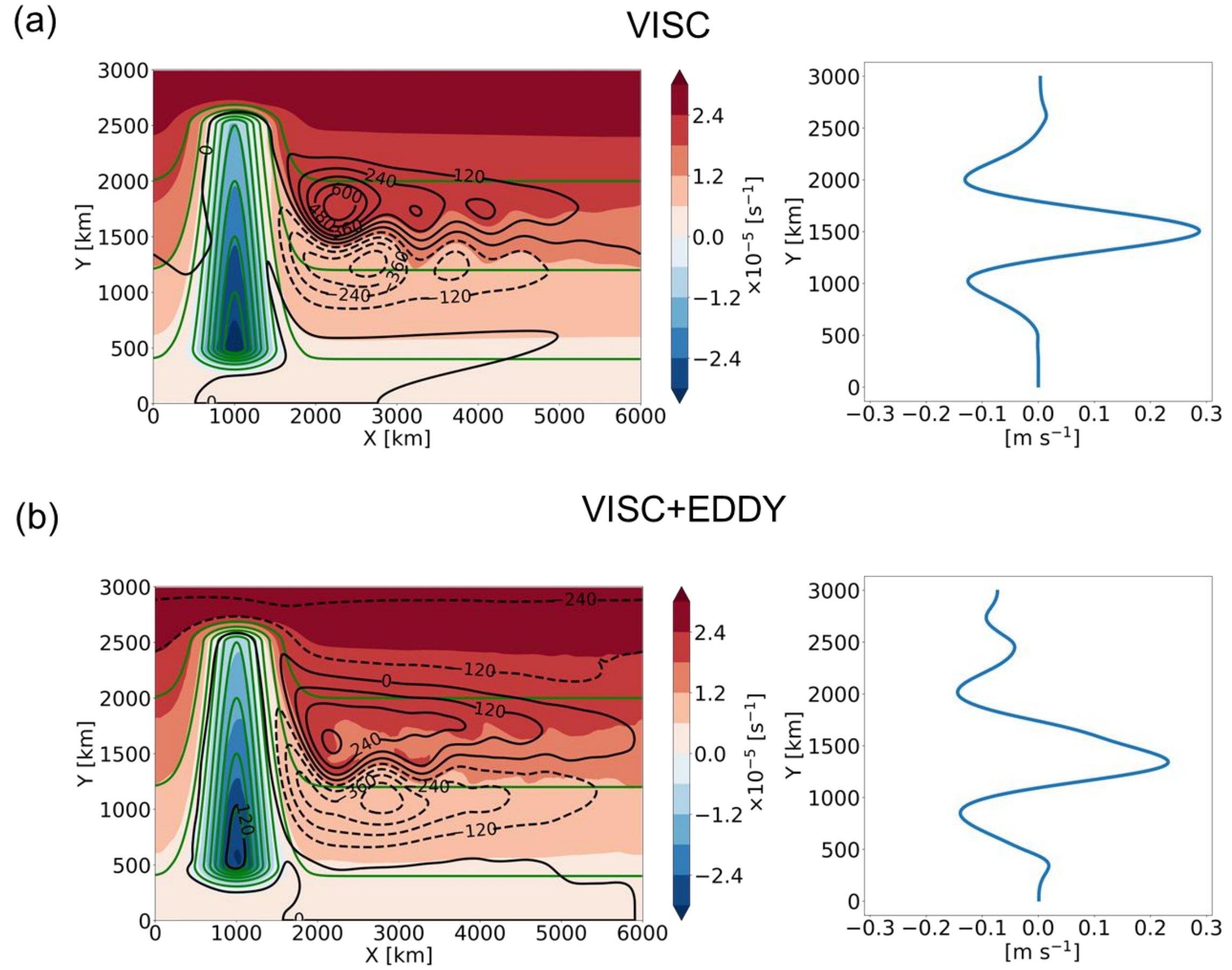

Figure 7 Color shades indicate the horizontal distribution of PV, and black contours indicate the streamfunctions (Left panel), and meridional profile of zonally-averaged zonal velocity (Right panel) in (a) VISC and (b) VISC+EDDY. The contours whose magnitude is larger than 600 Sv are not shown. Green contours denote the geostrophic contours with a contour interval of $2.0 \times 10^{-9}\ \mathrm{m}^{-1}\ \mathrm{s}^{-1}$.

## 4. Spinup of Westward Jet from the Viewpoint of Elliassen–Palm Flux

To elucidate the mechanism responsible for the generation of the westward circumpolar flow, spinup processes of the westward jet are analyzed based on the EP flux concept. The wave activity flux concept used in Section 3 is useful, but it can only be applied to time-averaged fields. Therefore, in the spinup analysis, the pseudo-momentum is defined in a zonally-averaged framework. The zonally-averaged momentum is linked with the pseudo-momentum by the relation

$$\frac{\partial [u]}{\partial t} = -\frac{\partial P}{\partial t} \approx \frac{\partial}{\partial y}(-[u^* v^*]), \tag{9}$$

where $P = [q^{*2}]/2[q_y]$ is the pseudo-momentum, $[\cdot]$ is the zonal averaging, and $\cdot^*$ is a deviation from the zonal mean (e.g., Vallis 2017). Although the transient component has been treated as *eddy* in the proceeding section, *eddy* is defined as the deviation from the zonal mean in this section. Therefore, the eddy term includes both contributions from Rossby waves and those arising from topographic effects such as western boundary currents. The righthand side corresponds to the divergence of the EP flux, $-[u^* v^*]\boldsymbol{j}$, where $\boldsymbol{j}$ is the meridional unit vector. The implication of the EP flux is same as that of the wave activity flux. If the EP flux is radiated from the region of disturbances, the pseudo-momentum density decreases in the region of disturbances. Therefore, Eq. (9) indicates that the Rossby wave radiations from disturbances correspond to an eastward acceleration in the region of the disturbances and a westward acceleration in the region of the EP flux convergence.

The time‑latitude plot of the EP flux divergence in LOW-DRAG is shown in Figure 8(a). During the first 100 days, weak positive and negative peaks are present between $y = 1400$ km and $y = 2700$ km. As will be shown later, a similar structure is formed in HIGH-DRAG as well, indicating that these structures are attributed to the advection by the western boundary current. During this period, the circumpolar transport is eastward (Figure 9(a)), and the meridional gradient of PV remains positive (Figure 10(a)), indicating that the eastward circumpolar current is stable to the barotropic instability. From the 100th day to the 300th day, small-scale structures begin to appear at approximately $y = 1500$ km . This signature corresponds to the onset of barotropic instability within the eastward circumpolar current. Figure 10(b) indicates that the eastward circumpolar current satisfies the necessary condition of barotropic instability. After the 150th day, the EP flux diverges between $y = 1000$ km and $y = 1500$ km, whereas it converges at the south of $y = 1000$ km and north of $y = 2000$ km. This signature corresponds to the wave activity flux radiation downstream of the eastward jet shown in Section 3. From around the same period, the transport begins to exhibit a pronounced decrease (Figure 9(a)), indicating that the westward momentum radiation from the unstable flow is essential for the westward circumpolar current formation.

For comparison, the case of HIGH-DRAG is also shown. Figure 8(b) illustrates that a development of the wind-driven circulation during the first 100 days is similar to that in the LOW-DRAG case shown in Figure 8(a). The circumpolar transport (Figure 9(b)), as in LOW-DRAG, increases during the first 50 days, then decreases until the 100th day. However, because the eastward circumpolar current remains stable throughout the entire period (Figure 10(b)), no Rossby wave radiation occurs; hence, the circumpolar transport becomes steady after 200 days.

These results indicate that the difference between the LOW-DRAG and HIGH-DRAG cases is caused by the meridional shear of the eastward circumpolar current in the spinup process. The smaller drag coefficient makes the eastward circumpolar transport more sheared, leading the barotropic instability.

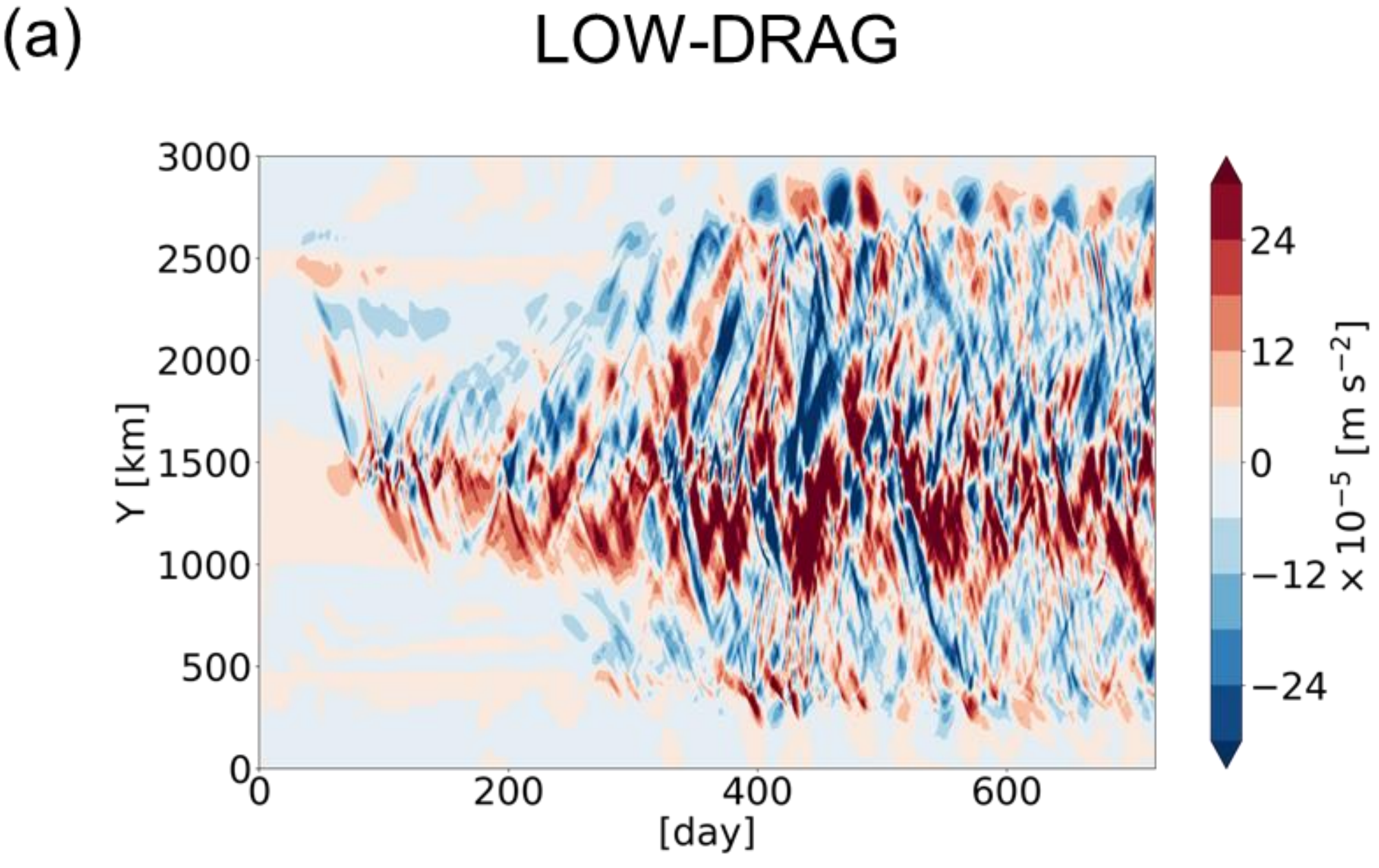

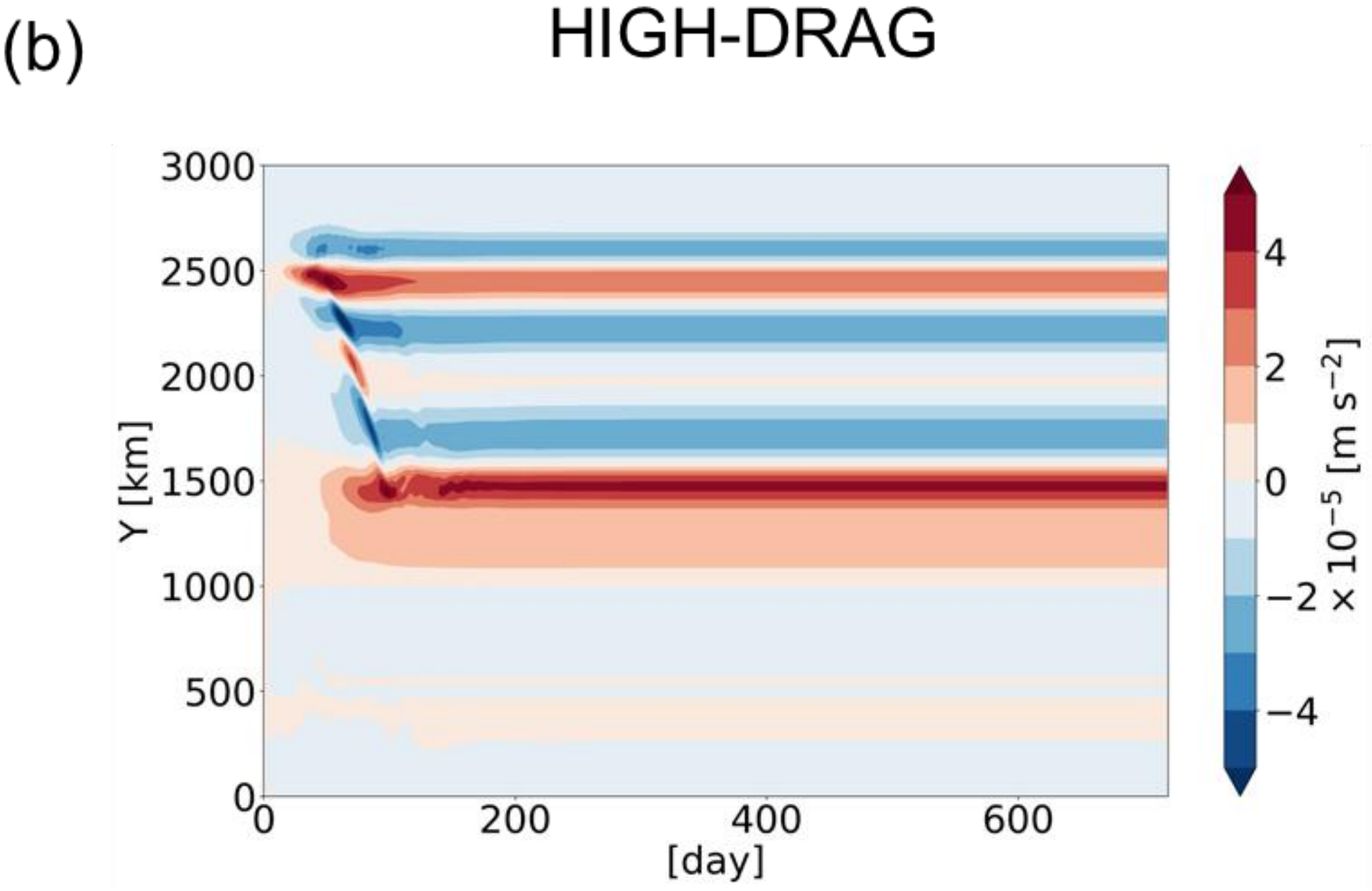

Figure 8 Time‑longitude plots of the EP flux divergence for (a) LOW-DRAG and (b) HIGH-DRAG. A positive value indicates an eastward acceleration and vice versa.

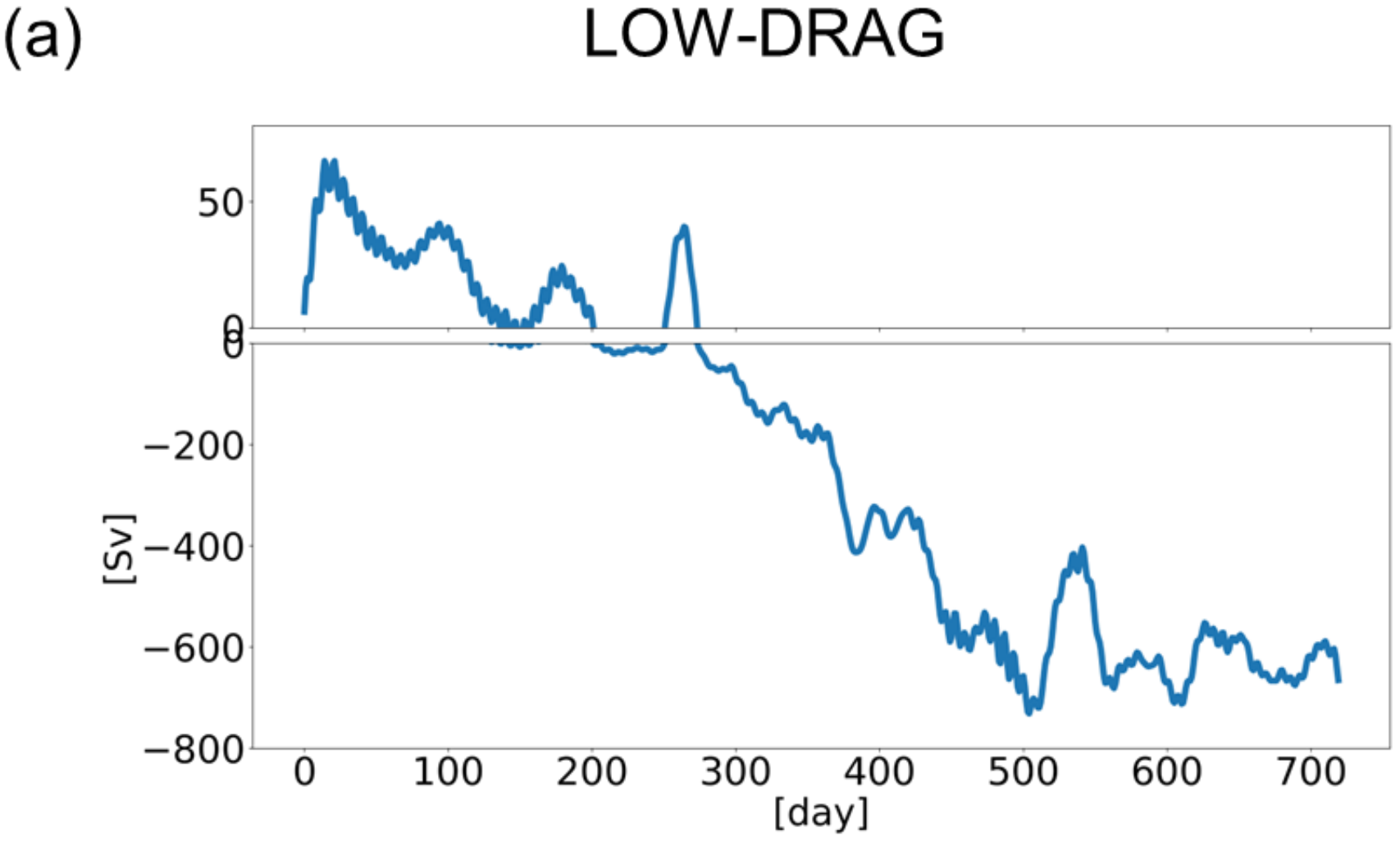

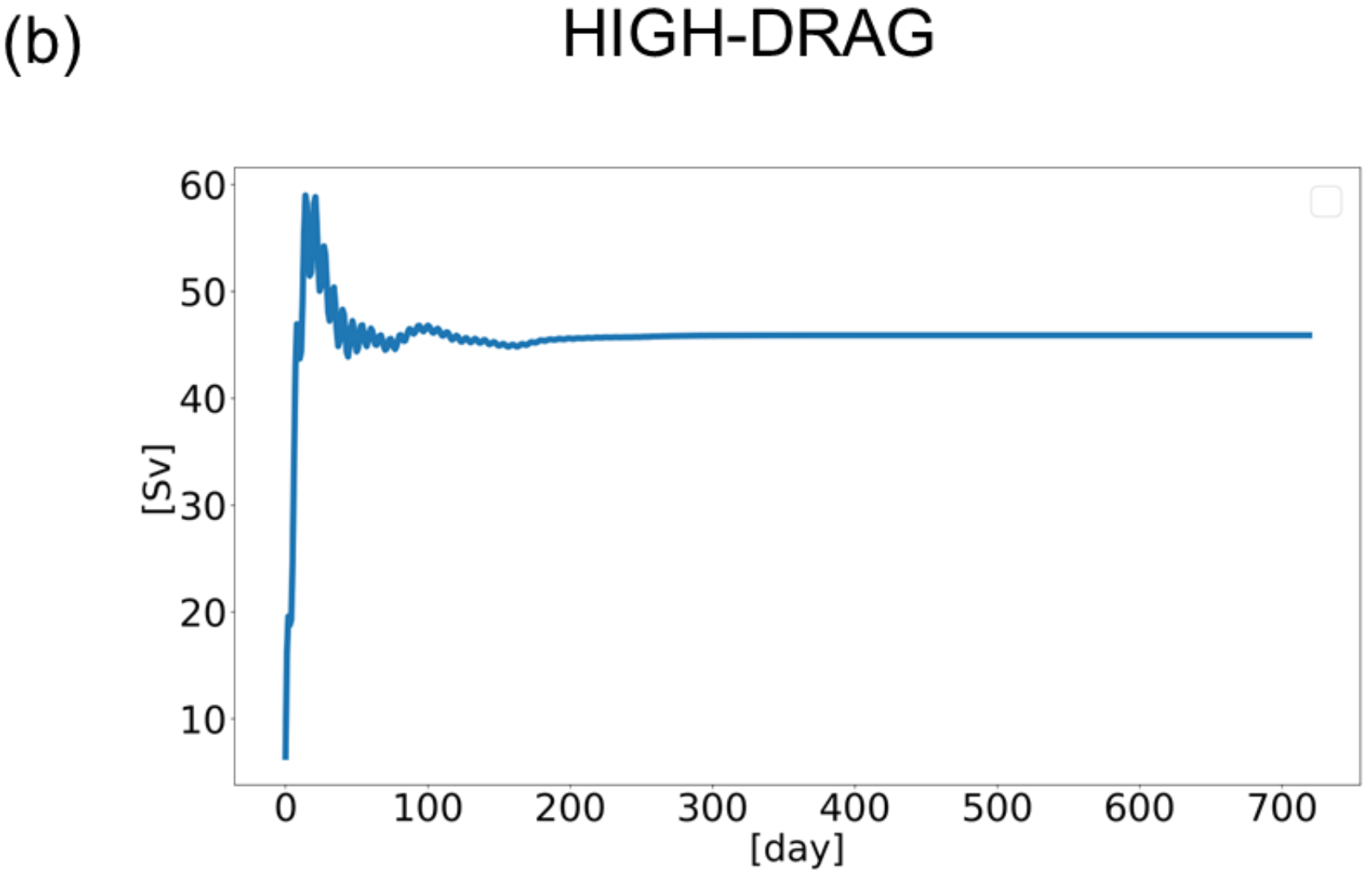

Figure 9 Time series of the circumpolar transport of (a) LOW-DRAG and (b) HIGH-DRAG. In the upper panel, the scale of the transport is adjusted separately for positive and negative values to emphasize changes in the eastward transport.

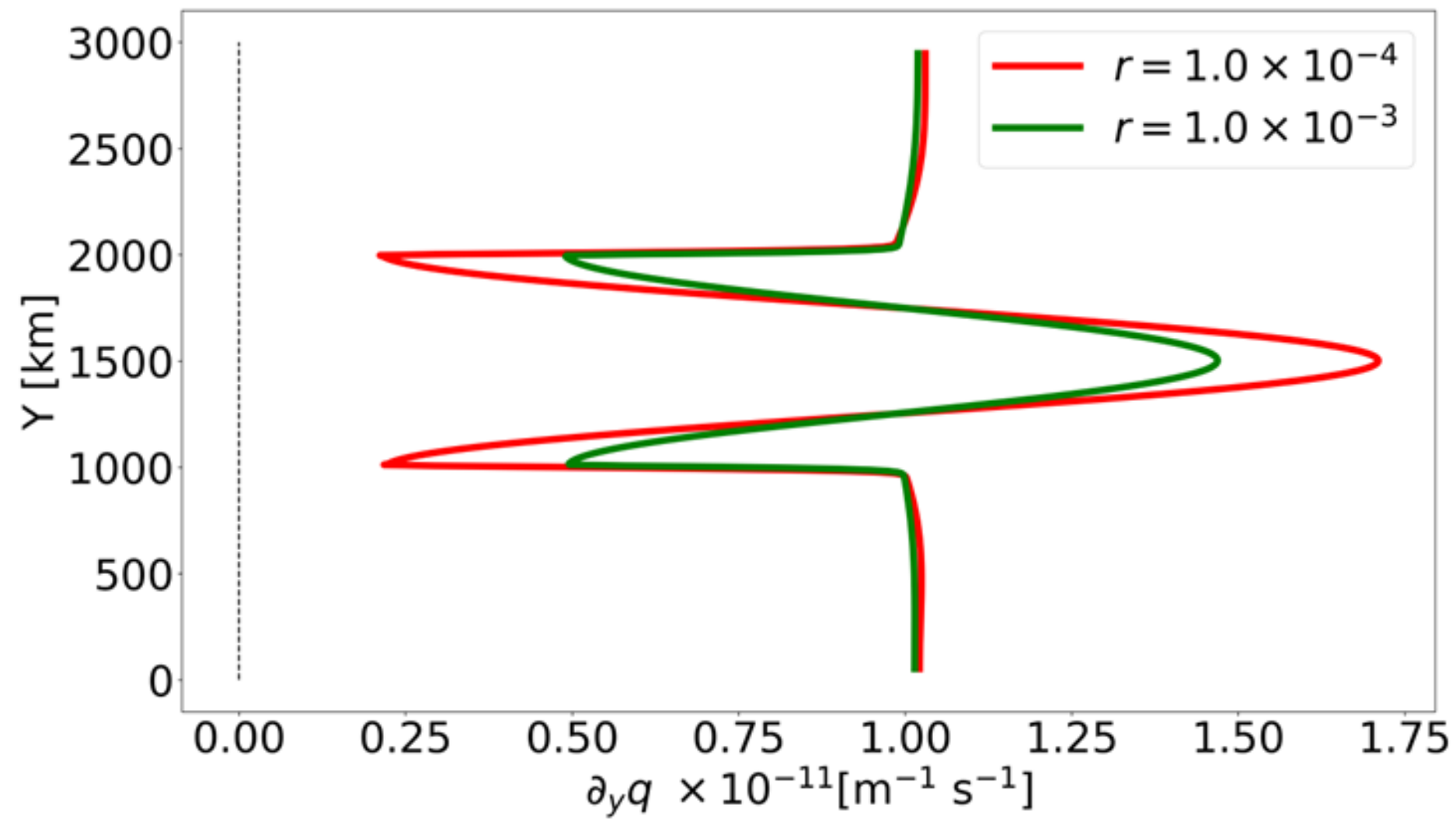

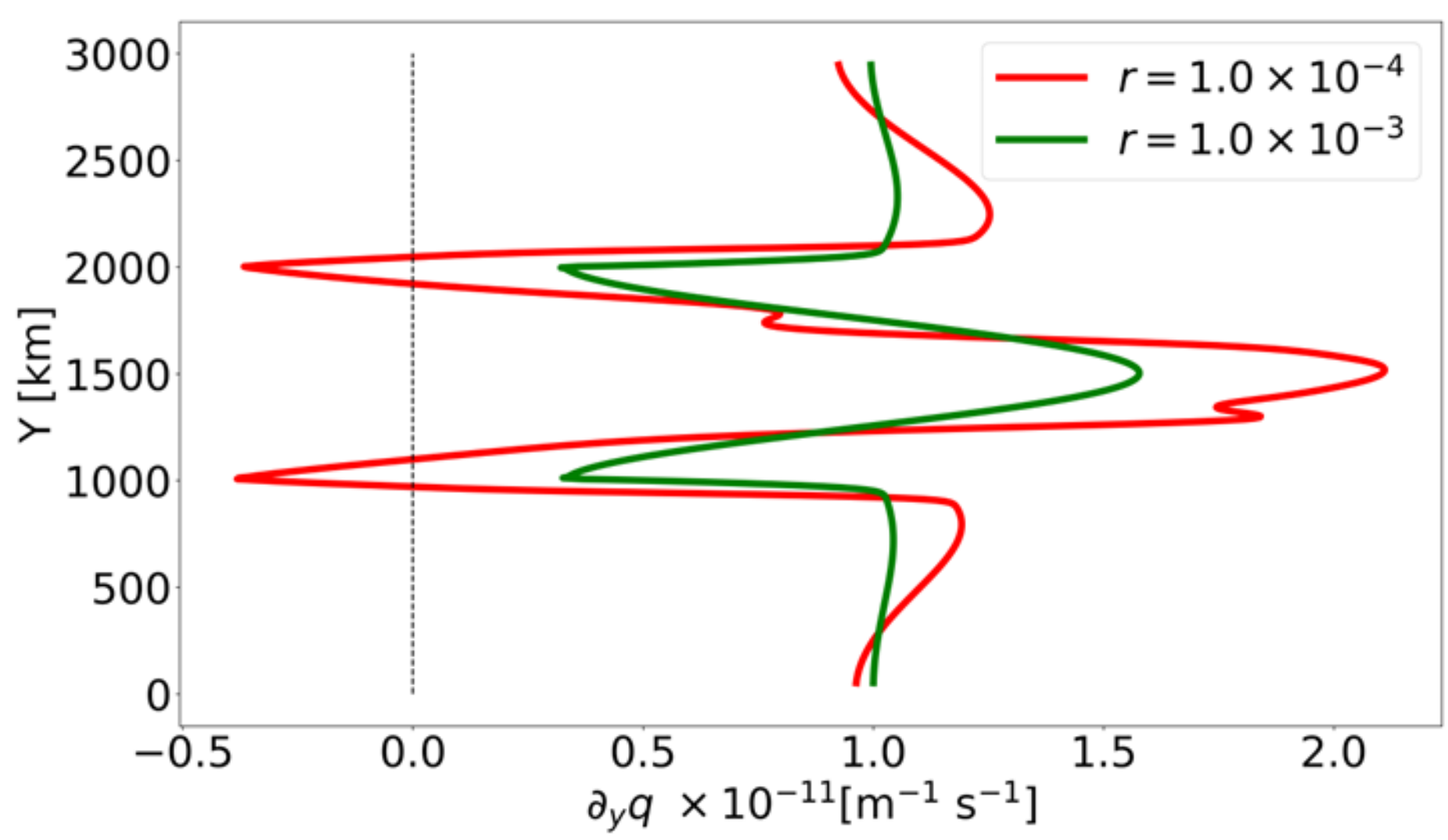

Figure 10 Meridional gradient of the PV along $x =$ 4000 km for LOW-DRAG (red) and HIGH-DRAG (green) on (a) 50th day and (b) 100th day. The dashed lines indicate the locations where the PV gradient becomes zero.

## 5. Robustness and limitations of our mechanism

Our experiments were conducted using high-amplitude bottom topography. Several studies (Wang and Huang 1995; Nadeau and Ferrari 2015) showed that barotropic dynamics in reentrant channels depend on the height of the bottom topography: when geostrophic contours are closed, basin-like dynamics emerge, as shown in this study, whereas when the contours are open, zonal jets dominate. To assess the robustness of the Rossby-wave-based frictional control mechanism identified in Section 3, we perform additional experiments with lower topographic height. Here, the topographic height is set to $\eta_0 = 300$ m and the other configurations are the same as LOW-DRAG and HIGH-DRAG, respectively. Figure 11(a) shows that the circumpolar transport is reversed in the low-drag case. The EKE exhibits large amplitudes within the eastward jet, indicating the presence of barotropic instability. According to the momentum analysis (not shown), the Reynolds stress redistributes westward momentum meridionally away from the center of the domain. Figure 11(b) demonstrates the circumpolar transport is eastward in the high-drag case. The EKE is approximately zero in this case. These features are identical to those in the high-topography case, indicating that our mechanism is not sensitive to the topographic height.

We also confirm that roles of Reynolds stress do not depend on the bottom drag. We conduct an additional experiment in which configurations are the same as HIGH-DRAG but the Reynolds stress diagnosed from LOW-DRAG is added to wind stress. A red contour in Figure 11(c) shows that westward circumpolar current is formed in the northern part of the domain. In Region N, westward forcing by the diagnosed Reynolds stress is mainly balanced by the bottom friction (not shown), which is the same as the LOW-DRAG case. This result suggests that the frictional control in our configurations regulates the transport by modifying the stability of the mean flow. Even in the high-drag case, if the Rossby wave radiation is sufficiently strong, the momentum redistribution occurs in a manner similar to that in the low-drag case.

Another frictional control mechanism has been reported in previous studies using reentrant barotropic channels with a monoscale topography (Uchimoto and Kubokawa 2005; Constantinou and Young 2017; Constantinou 2018). These studies showed that decreasing friction or viscosity results in the reduction in circumpolar transport and *eddy saturation*. Here, the eddy-saturated regime refers to a regime in which circumpolar transport becomes insensitive to the strength of the westerly wind forcing. Uchimoto and Kubokawa (2005) demonstrated that circumpolar transport is determined by a higher mode wave resonance. The reduction of viscosity enhances higher modes, which increases the form stress. The other studies (Constantinou and Young 2017; Constantinou 2018; Constantinou and Hogg 2019) suggested that a resonant wave is sustained by eddy flux. A reduction in friction enhances eddy activities and the associated eddy PV flux enhances the phase shift between standing waves and bottom topography. The smaller bottom drag therefore results in the increase in form stress.

The difference in the outcome of frictional control, *i.e.*, flow-reversal versus eddy saturation, arises from differences in the underlying bottom topography. In our configuration, the region downstream of the ridge is flat, and the flow behaves as a wind-driven gyre circulation. As shown in previous studies (e.g., Provost and Verron 1987; Waterman and Jayne 2011; Mizuta 2012; Waterman and Jayne 2012; Waterman and Hoskins 2013; Mizuta 2018a,b), the downstream region of an unstable wind-driven jet radiates Rossby waves through barotropic instability, which rectifies the jet and gyre. On the other hand, in configurations of previous studies, the zonal spacing of the bottom topography is narrow, so the flow is strongly constrained by topography. Eddies generated by barotropic instability therefore do not radiate Rossby waves but instead interact locally with standing waves. It is also noted that since bottom topography modifies the stability of the mean flow, barotropic instability occurs in their case even in the high-drag case. Because the Antarctic Circumpolar Current consists of regions that are strongly influenced by bottom topography and others that are not (Thompson and Naveira Garabato 2014; Wu et al. 2017; Matsuta 2022; Matsuta and Masumoto 2023), both mechanisms are worth being examined in future studies using stratified models.

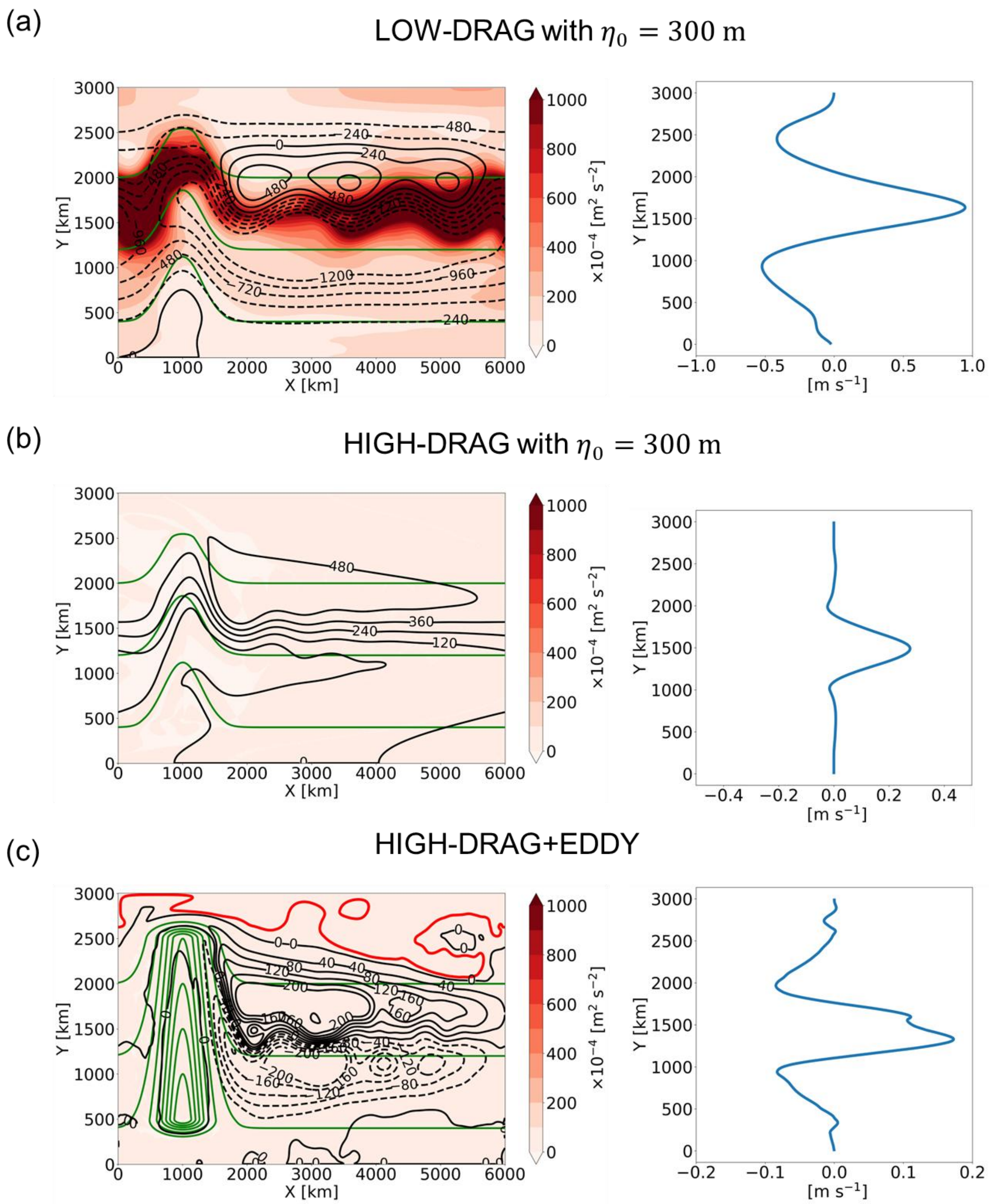

Figure 11 Horizontal distribution of EKE and streamfunctions (Left panel) and meridional profile of zonally-averaged zonal velocity (Right panel) for (a) LOW-DRAG, (b) HIGH-DRAG with $\eta_0 = 300$ m, and (c) HIGH-DRAG with eddy forcing. Color shades indicate the horizontal distribution of EKE, and black contours indicate the streamfunctions. The contours larger than 1200 Sv and 600 Sv are not shown in (a) and (b), respectively. A red contour in (c) indicates the −10 Sv contour. Green contours denote the geostrophic contours with a contour interval of $2.0 \times 10^{-9}$ m$^{-1}$ s$^{-1}$. The scales of horizontal axis of the right panel are different among the cases.

## 6. Summary and Conclusion

This study proposed a frictional control mechanism of barotropic reentrant channel models, the topography of which is sufficiently high for geostrophic contours to be locally blocked. The model was forced by steady westerly winds. In Section 3.1, it was demonstrated that the wind-driven eastward circumpolar transport is formed in the high-drag regime as shown in previous studies (Patmore et al. 2019; Matsuta and Mitsudera 2024). By contrast, in the low-drag regime, the westward circumpolar current is formed outside of the wind-driven gyre circulations (Figure 2). In this case, the eastward jet is in the gyre component rather than the circumpolar component; thus, it does not contribute to the net circumpolar transport. Reversing the net circumpolar transport was accompanied by eddy activities (Figure 3), suggesting that the barotropic instability sustains the westward circumpolar transport. The difference between the two regimes was also reflected in the meridional profile of the PV (Figure 4). In the low-drag regime, a PV staircase was formed, creating sharp eastward flows and relatively wide westward flows. At the flanks of the unstable eastward flow, the PV was homogenized, which is a signature of barotropic instability. In the high-drag regime, the meridional shear of the zonal flow was not strong enough to reverse the PV, making the planetary β-effect dominant. In fact, the flow was confirmed to reach a steady state. Furthermore, the behaviors in both regimes were confirmed to be unaffected by geometries of geostrophic contours as long as the equivalent western boundary was supported (Figure 2).

To investigate the mechanism sustaining the westward circumpolar current in the low-drag regime, the meridional profile of the momentum equation terms was investigated. A key difference between the low-drag regime and the high-drag regime is the eddy forcing. In the low-drag regime, the eddy forcing acts eastward in the unstable area, whereas it acts as the westward forcing outside the unstable eastward jet (Figure 5). On the other hand, eddy forcing was absent in the high-drag regime. Although the mean advection associated with the western boundary current also redistributed the westward momentum from the southern half of the domain to the northern half, the momentum redistribution by mean advection was common in the high-drag regime. To elucidate roles of these momentum redistributions for sustaining the westward circumpolar transport, momentum budget was investigated. In the low-drag regime, the westward eddy forcing was balanced by the bottom drag north of $y = 2000$ km and south of $y = 1000$ km, where westward circumpolar currents existed. In addition, the eastward eddy forcing in the unstable region was balanced by the intensified topographic form stress. These results indicated that the eddy forcing substantially altered the total momentum balance and sustained the westward net transport, although it was equals to zero in the total balance. On the other hand, the mean advection just altered the meridional profile of the topographic form stress and had little impacts on the total momentum balance.

The eddy forcing related to the Rossby wave radiation. Analyses of the wave activity flux (Figure 6) revealed that the Rossby wave was emitted from the high-EKE area, which removed the westward momentum from the region of disturbances. The wave activity flux converged in the regions of the westward circumpolar current, which corresponds to the effective westward forcing.

To confirm that the eddy forcing was essential for the westward circumpolar current formation, a strong-viscosity experiment was conducted, where the viscosity was high and other configurations were common to those of the low-drag experiment. In this case, the westward forcing associated with the mean advection was not changed from that in the low-drag case, whereas the eddy forcing was absent. Consequently, the circumpolar transport became steady eastward. When the diagnosed eddy forcing was added to the strong-viscosity model as an external forcing, the westward circumpolar transport was recovered. These results indicate that the eddy forcing is essential for the westward circumpolar current formation, whereas the momentum transport by the mean flow cannot drive the westward circumpolar current. A schematic of the suggested mechanism was shown in Figure 12.

Furthermore, the spinup processes in each case were analyzed using the EP flux framework (Figure 8). In the low-drag regime, the eastward circumpolar current was initially established (Figure 9). However, once the horizontal shear became sufficiently strong, the eastward jet satisfied the necessary condition for barotropic instability (Figure 10). Consequently, from around day 200 onward, Rossby waves were radiated, transporting westward momentum from the unstable jet to the region outside of disturbances (Figure 8). This momentum redistribution caused the net circumpolar transport to reverse its direction. These findings further support the conclusion that the Rossby waves radiation plays an essential role in the formation of the westward circumpolar current.

In Section 5, we discussed the robustness of our results. We showed that westward circumpolar transport is formed even when the topographic height is reduced to 300 m in the low-drag model. In addition, a westward flow also emerges when the Reynolds stress diagnosed from the low-drag case is imposed in the high-drag case. These results suggest that our proposed mechanism is valid if the Rossby wave radiation occurs.

Our results indicate that the momentum transport by the barotropic Rossby wave radiation possibly contributes to frictional control. Of course, since baroclinic instability is dominant in the ACC (e.g., Chen et al. 2014; Matsuta and Masumoto 2023; Matsuta et al. 2024), the frictional control associated with baroclinicity is unlikely to be negligible. However, if the bottom drag becomes small enough that barotropic instability cannot be ignored, the barotropic Rossby wave radiations may also contribute to the frictional control. Moreover, Rossby wave radiation may become non-negligible under strengthened westerlies because enhanced westerlies could lead to a non-negligible contribution from barotropic instability (Wu et al. 2017; Youngs et al. 2017). The barotropic model is highly idealized; thus, future studies must investigate whether a similar redistribution of westward momentum associated with Rossby waves occurs in stratified channel models.

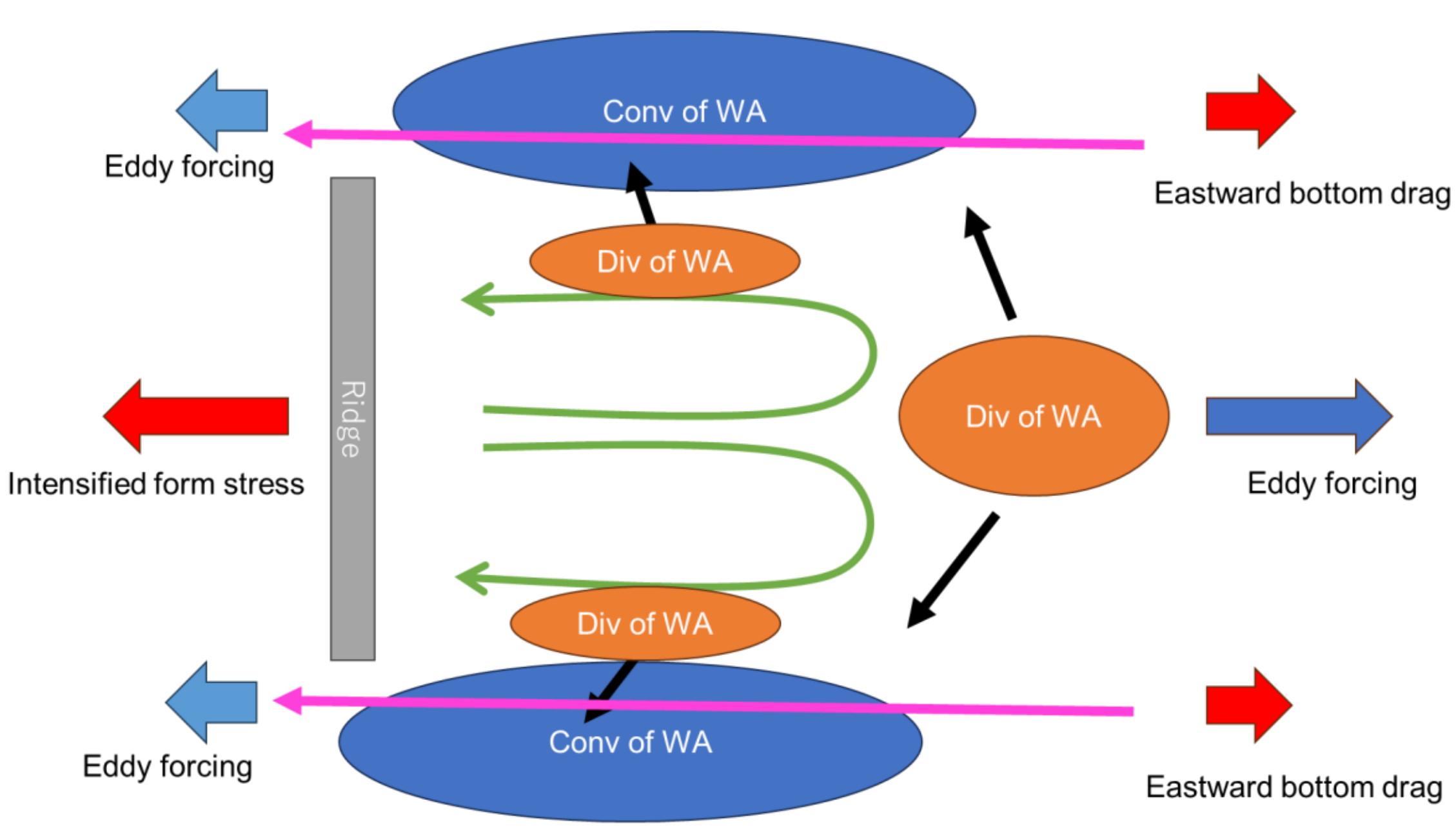

Figure 12 Schematic of the westward circumpolar current formation and maintenance mechanisms in the low-drag regime. The wave activity (black arrow) is radiated from the disturbances in the downstream region indicated by the orange circle. The wave activity is also radiated from the northern (southern) edge of the northern (southern) gyre. The divergence of the wave activity flux results in the eastward forcing at the latitude of the unstable jet, which increases the westward topographic form stress. The wave activity converges the north of the northern gyre and south of the southern gyre, forming and sustaining the westward circumpolar current. In these areas, the eastward bottom drag balances the eddy forcing.

## Appendix. Identification of the Rossby wave radiations based on the Hilbert Empirical Orthogonal Function

Since the background field exhibited zonal variations, identifying Rossby wave signals based solely on a simple dispersion relation was challenging. Accordingly, we employed the Hilbert Empirical Orthogonal Function (HEOF) analysis to identify the radiation of Rossby waves. An advantage of the HEOF is that the HEOF analysis identifies the propagating signals as modes from temporal and spatial data. In this work, the HEOF analysis is conducted using a Python library for EOF analyses, "xeofs" (Rieger and Levang 2024).

Before analyzing the HEOF results, we provided a brief introduction to the HEOF. A detailed information is found in previous papers (e.g., Barnett 1983; Hannachi et al. 2007; Matsuta et al. 2025). Let $\mathbf{q}_t = \{q(\boldsymbol{x}_k, t)\}^T{}_{k=1,\ldots,N_x}$ be the vector whose entry is a scaler value $q$ at time $t$ and at the position $\boldsymbol{x}_k$. Here, $N_x$ is the total number of grids and $\cdot^T$ indicates the transpose of a vector. A new complex field associated with $\mathbf{q}_t$ is defined as $\tilde{\mathbf{q}}_t = \mathbf{q}_t + i\mathcal{H}(\mathbf{q}_t)$, where $i$ is the imaginary unit and $\mathcal{H}(\cdot)$ indicates the Hilbert transformation. The real part of $\tilde{\mathbf{q}}_t$ is equal to $\mathbf{q}_t$. In the HEOF analysis, the new field is expanded as follows:

$$\tilde{\mathbf{q}}(\mathbf{x}_k, t) = \sum_{m=1}^{M} A_m(t) B_m^*(\mathbf{x}_k), \tag{A1}$$

where $A_m$ is the *m-th* time-dependent principal component given by

$$A_m(t) = \sum_{k=1}^{N_x} \tilde{\mathbf{q}}(\mathbf{x}_k, t) B_m(\boldsymbol{x}_k), \tag{A2}$$

and $B_m$ is the *m-th* eigenvector of the time-averaged Hermitian covariance matrix, *C*,

$$C = \frac{1}{N_t} \sum_{n=1}^{N_t} \tilde{\mathbf{q}}_{t_n}^* \tilde{\mathbf{q}}_{t_n}, \tag{A3}$$

where $\cdot^*$ indicates the adjoint of the complex matrix, $t_n$ is the time when $\tilde{\mathbf{q}}_t$ is observed, and $N_t$ is the length of the time series. The eigenvalue associated with $B_m$ corresponds to the fraction of total variance explained by the *m-th* mode.

From the HEOF modes, the spatial amplitude, $S_m$ and relative phase, $\theta_m$ of the *m-th* mode are obtained:

$$S_m(\mathbf{x}_k) = \left[B_m(\mathbf{x}_k) B_m^{*\,T}(\mathbf{x}_k)\right]^{\frac{1}{2}}, \tag{A4}$$

and

$$\theta_m(\mathbf{x}_k) = \arctan\left(\frac{\mathrm{Im}\,\left(B_m(\mathbf{x}_k)\right)}{\mathrm{Re}\,\left(B_m(\mathbf{x}_k)\right)}\right), \tag{A5}$$

where Re (·) and Im (·) return the real and imaginary parts of complex values, respectively. The spatial amplitude measures the intensity of each mode. The gradient of $\theta_m$ corresponds to the local wave number if the mode is represented as a single plane wave. Similarly, the temporal amplitude $R_m$ and phase $\phi_m$ are represented as

$$R_m(t) = [A_m(t) A_m^{*}\,(t)]^{\frac{1}{2}}, \tag{A6}$$

and

$$\phi_m(t) = \arctan\left(\frac{\mathrm{Im}\,(A_m(t))}{\mathrm{Re}\,(A_m(t))}\right). \tag{A7}$$

The HEOF analysis is conducted for the daily sea surface elevation data in the period from 10th to 11th year for the LOW-DRAG experiment. The first mode explains 30 % of the variations. Figure A1(a) shows that the first model reflects the active eddy variations downstream of x=3000 km (Figure 6(a)). Figure A1(b) demonstrates that the phase lines extend northwestward in the northern half of Box A, corresponding to a southwestward wavenumber vector. If the dispersion relation for barotropic Rossby waves is assumed, the southwestward wavenumber vectors correspond to the northward group velocity and negative value of $\overline{u'v'}$. By contrast, in the southern half of Box A, the phase lines extend toward the southwest, which corresponds to the southward wave propagation and positive $\overline{u'v'}$. Indeed, color shades of Figure A1(b) indicate that $\overline{u'v'}$ is negative in the northern part of Box A and the opposite is true in the southern part. Although less pronounced than in Box A, in Box B (Box C), the phase line is northwestward (southwestward), which accompanies the negative (positive) value of $\overline{u'v'}$. These results support the idea that the wave activity flux analyzed in Section 3.3 (Figure 6) corresponds to the radiation of Rossby waves.

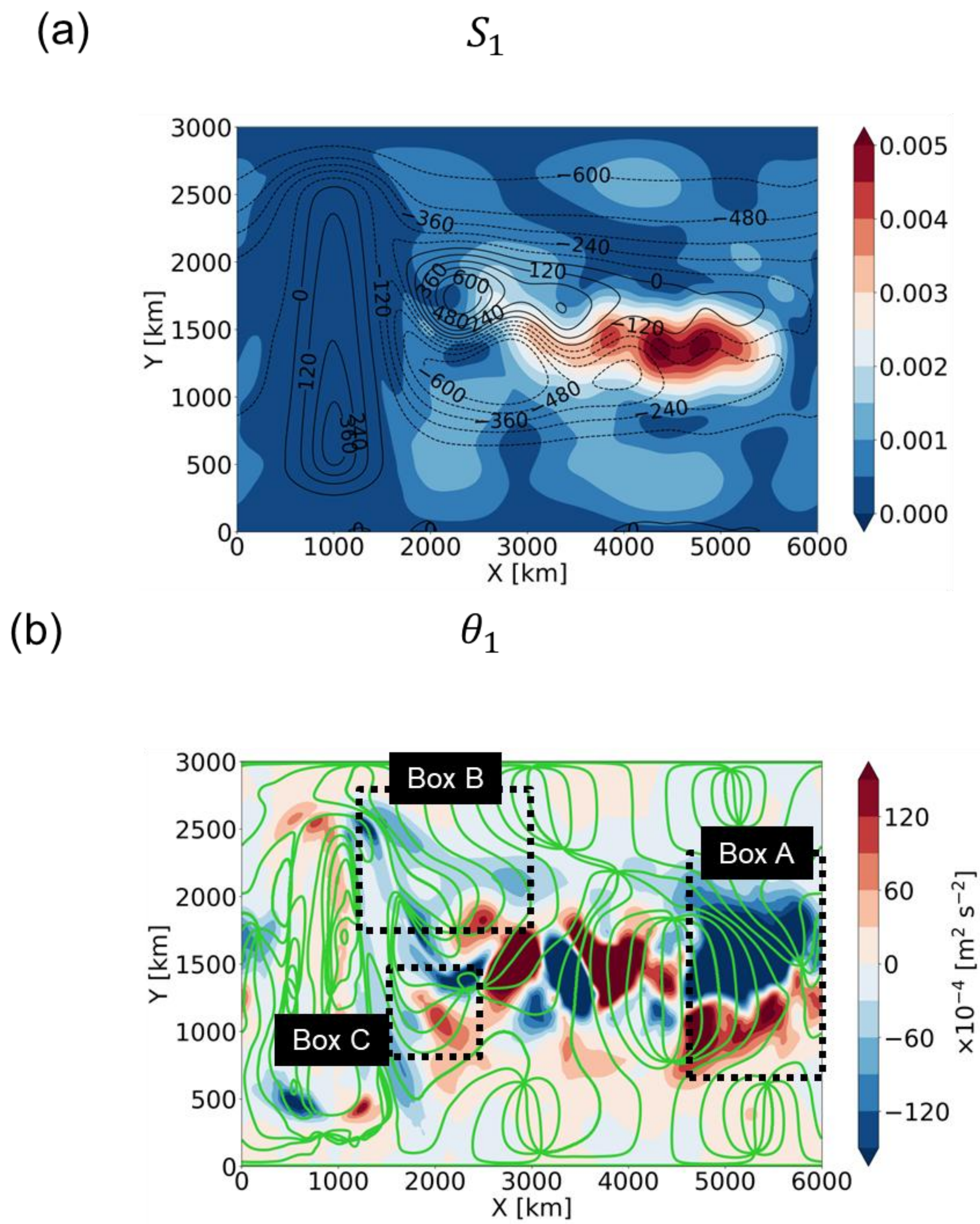

Figure A1 Results of the HEOF analysis for the sea surface elevation in the period from 10th to 11th year of LOW-DRAG. (a) Color shades indicate the horizontal distribution of the spatial amplitude for the first mode of HEOF. Black contours indicate the streamfunctions whose contour interval is 120 Sv. (b) Color shades indicate the horizontal distribution of $\overline{u'v'}$. Lime green contours indicate the relative phase of the HEOF first mode. The contour interval is $\pi/4$. Dashed rectangular boxes explained in the text.

## Acknowledgements

Takuro Matsuta was supported by JSPS Grant-in-Aid for JSPS Fellows No. 22J00651 and JSPS Grant-in-Aid No. 24K17120. Numerical experiments were conducted using the Earth simulator (ES4) at Japan Agency for Marine-Earth Science and Technology.

## Data availability statement.

MITgcm is available at https://mitgcm.readthedocs.io/en/latest/. Configuration files for our experiments are available at doi:10.5281/zenodo.18265192.